# Abundance, Major Element Composition and Size of Components and Matrix in CV, CO and Acfer 094 Chondrites

Denton S. EBEL[1,2,3*], Chelsea BRUNNER[4], Kevin KONRAD[5], Kristin LEFTWICH[6], Isabelle ERB[7], Muzhou LU[8], Hugo RODRIGUEZ[9], Ellen J. CRAPSTER-PREGONT[1,2], Jon M. FRIEDRICH[1,10], and Michael K. WEISBERG[1,3,11]

[1]Department of Earth and Planetary Sciences, American Museum of Natural History, NY, NY 10024,

[2]Lamont-Doherty Earth Observatory, Columbia University, Palisades, New York 10964, USA,

[3]Graduate Center of the City University of New York, New York

[4]Barrick Gold, Nevada (cebrunner@gmail.com),

[5]College of Earth, Ocean, and Atmospheric Sciences. Oregon State University, Corvallis, OR, USA (konradke@geo.oregonstate.edu),

[6]Dept. of Geological Sciences, Indiana University, Bloomington IN, USA (k.m.leftwich@gmail.com),

[7]Yale University, 409 Prospect St, New Haven, CT 06511, USA (isabellerb@gmail.com),

[8]Department of Astronomy and Physics, Williams College, MA, USA (muzhoulu@yahoo.com),

[9]SandRidge Energy, Houston, TX, USA (hugor@miners.utep.edu),

[10]Department of Chemistry, Fordham University, 441 East Fordham Road, Bronx, NY 10458, USA,

[11]Department of Physical Sciences, Kingsborough College, 2001 Oriental Blvd., Brooklyn, New York 11235, USA

* Corresponding author: debel@amnh.org

**Abstract** -- The relative abundances and chemical compositions of the macroscopic components or "inclusions" (chondrules and refractory inclusions) and fine-grained mineral matrix in chondritic meteorites provide constraints on astrophysical theories of inclusion formation and chondrite accretion. We present new techniques for analysis of low count/pixel Si, Mg, Ca, Al, Ti and Fe x-ray intensity maps of rock sections, and apply them to large areas of CO and CV chondrites, and the ungrouped Acfer 094 chondrite. For many thousands of manually segmented and type-identified inclusions, we are able to assess, pixel-by-pixel, the major element content of each inclusion. We quantify the total fraction of those elements accounted for by various types of inclusion and matrix. Among CO chondrites, both matrix and inclusion Mg/Si ratios approach the solar (and bulk CO) ratio with increasing petrologic grade, but Si remains enriched in inclusions relative to matrix. The oxidized CV chondrites with higher matrix/inclusion ratios exhibit more severe aqueous alteration (oxidation), and their excess matrix accounts for their higher porosity relative to reduced CV chondrites. Porosity could accommodate an original ice component of matrix as the direct cause of local alteration of oxidized CV chondrites. We confirm that major element abundances among inclusions differ greatly, across a wide range of CO and CV chondrites. These abundances in all cases add up to near-chondritic (solar) bulk abundance ratios in these chondrites, despite wide variations in matrix/inclusion ratios and inclusion sizes: chondrite components are complementary. This complementarity provides a robust meteoritic constraint for astrophysical disk models.

## 1. INTRODUCTION

Chondritic meteorites are extremely diverse, yet the first order observation is that they are all "chondritic": they contain near-solar ratios of major and trace refractory elements (e.g., Wood, 1963). They are divided into groups that contain a wide variety of inclusions such as chondrules, Ca-, Al-rich inclusions (CAIs), amoeboid olivine aggregates (AOAs), and others (clasts, in a sedimentary sense). The size distributions and abundances of the various inclusions are remarkably characteristic of each chondrite group (Van Schmus and Wood, 1967; Van Schmus and Hayes, 1974; Weisberg et al., 2006). These variations are thought to represent localized heterogeneity of the nebular environments in which chondrites formed (Scott and Taylor, 1983; Brearley and Jones, 1998; Jones and Schilk, 2009; Jones, 2012). These gross textural features are consistent with classification based on oxygen isotopic analysis (Clayton et al. 1976) and bulk chemical data (e.g., Wasson and Kallemeyn, 1988; Nittler et al., 2004). For example, the Vigarano-type (CV) carbonaceous chondrites were distinguished by Van Schmus (1969) primarily by textural criteria: chondrules are large and contain opaque mineral grains. The most cited measurements of inclusion and matrix abundances in CO and CV chondrites are those of McSween (1977a, b). In order to address the major unanswered question (MacPherson et al., 2005, p.242): "Why are CAIs and chondrules sorted into different size and type populations within different chondrite types", it is necessary to document with some quantitative accuracy those actual sizes and type abundances in the chondrites. How well are these inclusions sorted, and what is the budget of major elements among them? Here, we present new data on the size, abundance, and composition of inclusions and matrix in large areas of CO and CV chondrites.

Chondritic meteorites from asteroid belt sources exhibit a paradoxical mixture of hot and cold components. Their matrices preserve heat-sensitive, "cold" components including presolar carbide and graphite grains (Mendybaev et al., 2002; Huss et al., 1981, 2003, 2006) and macromolecular carbon (Alexander et al., 2007) that could not survive above ~700 K. Yet many inclusions (chondrules, CAIs) experienced temperatures above 1500 K prior to accretion (Ebel, 2006). Even Kuiper belt (~40 AU) comet Wild 2 contains chondrule-like (Nakamura et al., 2008), and CAI-like high-temperature inclusions (Simon et al., 2008). The relative abundances of inclusions, their bulk chemical characteristics, and their abundances relative to matrix determine the bulk composition of every class of chondrite.

The accretion of chondrule precursor material is presumably reflected in the sizes and chemical compositions of inclusions found in chondrites (Rubin, 2000; Jones, 2012). Here, we document how major elements Si, Mg, Fe, Al, Ca and Ti are distributed among inclusions and matrix in CO and CV chondrites. We discuss the constraints that the compositions of inclusions and matrix components, and sizes and abundances of inclusions, might place on astrophysical mechanisms for inclusion formation and chondrite parent body accretion.

We measure the relative proportions of inclusions of different bulk chemical composition and texture (e.g., CAIs, chondrules, AOAs), and inclusion/matrix ratios, by application of modern methods of aggregate clast (inclusion) analysis (Ebel et al., 2008a), building digitally upon the pioneering petrographic work of McSween (1977a, b). We analyze element-specific x-ray intensity maps of large slab areas of Allende, polished sections of Allende and seven other CV chondrites, and large polished section areas of six CO chondrites and Acfer 094 (C2 ungrouped). These results are comparable to but differ from previous optical point-counting analyses because the numbers of points are large, and compositional data is recovered.

We then use the ability to query each pixel in each inclusion to extract x-ray intensity statistics for the major elements in each inclusion, and also in matrix. This data may be "mined" in a variety of ways. We use it here to assess the chemical variability among clast types and matrix. We discuss these findings in the context of contributions of clast and matrix to the chondritic elemental abundances in CV and CO chondrites. In analyzing thousands of inclusions across a wide range of meteorites in these classes, we have endeavored to measure the overall characteristics of the "forest" of inclusions, although we also discovered many individual "trees" that appear compelling for future investigation.

## 2. PREVIOUS WORK

### 2.1. Modal abundances

Detailed determination of the modal abundances and sizes of chondritic components in CO and CV chondrites began in earnest with the thin section point-counting of McSween (1977a, b) and the inclusion analysis of King and King (1978), following Dodd (1976; cf. Chayes, 1956; Higgins, 2006). McSween (1977a, b) noted that "more accurate modes could be obtained from large slabs". King and King (1978) counted only objects with apparent diameter > 0.1mm. The matrix and clast abundances reported by McSween (1977a, b) have reappeared in a great many review articles and papers (e.g., Rubin, 2010). Grossman and Brearley (2005) measured matrix/inclusion/metal/sulfide ratios by image analysis of small x-ray mapped regions of CO and ordinary chondrites, including maps of Na, S, and K. Hezel et al. (2008) reviewed previous work, and used previous and new measurements to estimate CAI abundances in carbonaceous chondrites using Poisson statistical methods. Fisher et al. (2014) have adopted digital methods similar to those described here to study large slabs of Allende (CV).

The classification of objects into categories using optical microscopy is challenging. McSween (1977a, b) identified chondrules (spherical, once molten), inclusions, and lithic and mineral fragments, with subdivisions including amoeboid olivine aggregates (AOAs) and non-igneous CAIs (see Electronic Annex, Table S1). Several classification schemes have been proposed for classes of objects in chondrites (e.g., Grossman, 1980; Kornacki et al., 1983; Cohen et al., 1983). Another complicating factor is the ubiquitous rims on both chondrules and CAIs (Rubin and Wasson, 1987; Metzler et al., 1992; Metzler and Bischoff, 1996; Cosarinsky et al., 2008; Huss et al., 2005; Rubin, 2010) and distinguishing rim/matrix boundaries.

### 2.2. Inclusion sizes in CV and CO chondrites

King and King (1978) examined grain size distributions in CV and CO chondrite thin sections. They found that the CV chondrite chondrules and fragments were distinctly coarser grained than those in CO chondrites. Paque and Cuzzi (1997) separated chondrules from Allende and ALH 84028 (CV) and ALH 85033 (L4) using freeze/thaw disaggregation (MacPherson et al., 1980), and determined their sizes, masses, and rim thicknesses. They determined that the calculated rim volume determined from 2D photomicrograph observations was approximately equal to the volume of the underlying chondrule. Mean chondrule diameters for Allende (850 µm), ALH 84028 (970 µm) and ALH 85033 (720 µm) were consistent with earlier work (Dodd, 1976; Grossman et al., 1988). However, in later work they (Teitler et al., 2010) cautioned that some bias may be present in the dataset. Rubin (2010) found a mean chondrule rim thickness of 400 µm in CV chondrites but noted that the median value is likely closer to 200-250 µm (Rubin, 1984). Those measurements included no attempt to correct for stereological 2D to 3D correction to represent true volumetric size distributions. However, Hezel (2007) developed a method to correct 2D size data for CV chondrite inclusions. Rubin et al.

(1985) and Rubin (1989; 2000) measured chondrule sizes in CO3 thin sections. They found a mean size of 150 µm, with only slight variation between different individual CO chondrites. Rubin (1989) also examined the size frequency distributions of chondrules by textural type. Eisenhour (1996) developed and applied stereological 2D to 3D corrections to Rubin's (1989) results. A complete review of chondrule size determinations across all chondrule-bearing meteorites is presented by Friedrich et al. (2014).

## 2.3. Bulk compositions

McSween (1977a, b, c) measured the bulk compositions of inclusions using broad beam electron probe microanalysis (EPMA) on thin sections. He noted a large spread in Ca and Al compositions among CAIs in carbonaceous chondrites (McSween, 1977c). McSween and Richardson (1977) reported depletion (enrichment) factors (element/Si[matrix])/(element/Si[bulk]) for a wide range of carbonaceous chondrites. Jones et al. (2005) review previous measurements of chondrule bulk composition. Jones and Schilk (2009) reported the bulk compositions of 91 chondrules from Mokoia (CV) obtained by instrumental neutron activation analysis (INAA). Hezel and Palme (2008, 2010) reported chondrule compositions modally reconstructed from electron probe data for CO and CV chondrites. Methods are compared under "Sample Mapping", below.

Chondrites in bulk have "chondritic" (solar) ratios of major non-volatile elements within narrow bounds (Table 1). Wasson and Kallemeyn (1988), Jarosewich (1990) and Wolf and Palme (2001) report mean bulk compositions for CO and CV chondrites (cf., Nittler et al., 2004). The lithophile elements Si, Mg, Fe, Ti, Al, and Ca vary only slightly in absolute abundance among the CV and CO chondrites (Larimer and Wasson, 1988; Table 1). The CV are slightly depleted in Fe relative to CO (by ~5%), and enriched in Al, Ca and Ti (by 10 - 20%), with Wolf and Palme (2001) finding smaller differences between CV and CO than did Wasson and Kallemeyn (1988). The CO and CV have CI chondritic Si/Mg ratios (Wolf and Palme, 2001; Table 1). It has been shown that even the altered (>3.6) CV3 Allende is highly homogeneous, and chondritic, in Si, Mg and Fe at 0.6g (~150 mm$^3$) sampling resolution (Stracke et al., 2012). It has been known for a long time that the matrices of carbonaceous chondrites differ from chondrite bulk compositions. McSween and Richardson (1977) compared matrix Mg/Si (0.822 ± 0.053) with bulk (~0.910), and found Ti/Al of 0.86 in CV and 0.57 in CO matrix, albeit of "questionable significance" (their Table 4).

**Table 1:** Major refractory elements in CV and CO chondrites. Atom fraction data normalized to Mg and the CI value reported from each reference are from Wolf and Palme (2001, WP01) and Wasson and Kallemeyn (1988, WK88).

|           | Si    | Fe    | Ca    | Al    | Ti    | Ti/Al | Ca/Al |
|-----------|-------|-------|-------|-------|-------|-------|-------|
| CO (WK88) | 1.013 | 0.912 | 1.149 | 1.112 | 1.242 | 1.117 | 1.033 |
| CO (WP01) | 1.008 | 0.910 | 1.199 | 1.160 | 1.136 | 0.979 | 1.034 |
| CV (WK88) | 0.994 | 0.864 | 1.382 | 1.361 | 1.561 | 1.147 | 1.015 |
| CV (WP01) | 0.969 | 0.804 | 1.315 | 1.305 | 1.262 | 0.968 | 1.008 |

## 3. METHODS

## 3.1. Samples

Polished slabs (not epoxied), polished epoxide thick sections (PS), and polished thin sections (PTS) were selected from the AMNH meteorite collection (Table 2). Allende (CV3), of the oxidized Allende-like subgroup (Weisberg et al., 1997) was analyzed in great detail (Brunner et al., 2008). Also analyzed were a single section each of CV oxidized-A subgroup Tibooburra (Fitzgerald and Jaques, 1982), oxidized Bali-like subgroup Bali and Mokoia, and reduced subgroup Vigarano and Leoville, and Nova 002 (not previously assigned to a subgroup). Sections of CO chondrites include Colony, Kainsaz, Ornans, Lancé and Warrenton. We also studied the ungrouped C2 (CO-like) chondrite Acfer 094 (Newton et al., 1995; Krot et al., 2004a; Grossman and Brearley, 2005).

**Table 2:** CV and CO chondrites studied. CV types B06 (Bonal et al., 2006) are based on Raman spectroscopy, and G95 (Guimon et al., 1995) are based on thermoluminescence sensitivity. CV subclasses are from Weisberg et al. (1997). CO types are from SJ90 (Scott and Jones 1990), GB5 (Grossman and Brearley, 2005), and M77 (McSween, 1977b) by mineral chemistry and S91 (Sears et al. 1991, their Table 4) by thermoluminescence. Tibooburra, Nova 002 and Acfer 094 have not been characterized by petrologic grade. Shock stages of CV are from Scott et al. (1992), who note that Vigarano is a breccia. All the CO are shock stage S1 (Grossman and Brearley, 2005). Acfer 094 is an ungrouped C2 chondrite, shock stage S1 (Newton et al., 1995). Porosity data are from Macke et al. (2011). *Shock stage for Nova 002 was determined by M. Weisberg for this work.

| CV | subclass | fall/find | type(G95) | type(B06) | shock | porosity |
|---|---|---|---|---|---|---|
| Allende | OxA | fall-1969 | 3.2 | >3.6 | S1 | 21.9 |
| Tibooburra | OxA | find | n.d. | n.d. | n.d. | n.d. |
| Nova 002 | n.d. | find | n.d. | n.d. | S1* | 6.0 |
| Bali | OxB | fall-1907 | 3.0 | ~3.6 | S3 | n.d. |
| Mokoia | OxB | fall-1908 | 3.2 | ~3.6 | S1 | 27.7 |
| Vigarano | Red | fall-1910 | 3.3 | 3.1-3.4 | S1-S2 | 8.3 |
| Leoville | Red | find | 3.0 | 3.1-3.4 | S3 | 4.84 |
| Grosnaja | OxB | fall-1861 | 3.0 | ~3.6 | S3 | n.d. |

| CO | fall/find | type(SJ90) | type(S91) | type(GB5) | (M77) | porosity |
|---|---|---|---|---|---|---|
| Colony | find | 3.0 | 3.0 | 3.0 | n.d. | 9.2 |
| Kainsaz | fall-1937 | 3.1 | 3.2 | 3.2 | I | 9.3 |
| Ornans | fall-1868 | 3.3 | 3.4 | 3.4 | II | 34.2 |
| Lancé | fall-1872 | 3.4 | 3.4 | 3.5 | II | 9.2 |
| Warrenton | fall-1877 | 3.6 | 3.6 | 3.7 | III | 24 |
| Acfer 094 | find | n.d. | n.d. | C2-ung | | n.d. |

*3.1.1. Alteration*

The CV3 chondrites are all affected by secondary processes (Krot et al., 1995; Kimura and Ikeda, 1998; Huss et al., 2003, 2006). Matrix of the oxidized subgroup contains magnetite and not low-Ni metal (Brearley and Jones, 1998, p.221). Grosnaja (AMNH PTS 455-1) was mapped (8 μm/pxl, $6.35 \times 10^6$ pixels) for this project, but appeared sufficiently altered that inclusions could not be distinguished clearly from matrix. However, Allende has been determined by many to be the most altered of our suite of CV chondrites (Ma and Krot, 2014). Bonal et al. (2006) determined an "unambiguous" hierarchy in thermal metamorphic grade based on organic maturation of Grosnaja < Mokoia ~ Bali < Allende (Table 2). Zolensky et al. (1993) concluded that CV chondrites experienced alteration at high water/rock ratios. Krot et al. (1995) found that matrix olivine becomes more equilibrated in the order Mokoia < Vigarano < Grosnaja < Allende. They concluded that the reason for this may be increasing peak metamorphic temperatures, or different nebula conditions. Krot et al. (1995) note that Grosnaja, Mokoia and Bali contain considerably more phyllosilicates than Allende, and observe coarser-grained matrix olivine in oxidized CV than in reduced CV chondrites. Hydrous alteration in Bali and Mokoia varies among samples, with Bali generally the most extensively altered (Keller et al., 1994), brecciated, and with "extreme heterogeneity of alteration" (Scott et al., 1997). Krot et al. (1995) concluded that there is strong evidence that Bali, Grosnaja and Mokoia were aqueously altered on their parent body, based on similar inferred temperatures and sulfur and oxygen fugacities, and that Allende experienced significant thermal processing (Scott and Krot, 2003). Nearly all metal in Allende has been oxidized or sulfidized. Among the reduced CV, presolar SiC and diamond abundances indicate Leoville is of lower petrologic grade than Vigarano (Huss et al., 2006), consistent with the thermoluminescence sequence (Table 2; Guimon et al., 1995).

The CO3 chondrites, by contrast, exhibit a more well-established metamorphic sequence (McSween, 1977b; Chizmadia et al., 2002; Grossman and Brearley, 2005; Huss et al., 2006). Chizmadia et al. (2002) noted that AOA in CO3.0 are angular or fragmented, but in subtypes ≥ 3.4 they are more rounded and embayed, and matrix olivine is completely equilibrated (Brearley and Jones, 1998). The abundance of opaque metal and troilite declines in subtypes ≥ 3.7, and the fayalite content of olivine in AOAs increases with petrologic grade (Chizmadia et al., 2002). But even in CO 3.7, AOAs and matrix have not equilibrated with each other. They only show signs of approaching equilibrium (Huss et al., 2006). Here, we present and interpret aspects of the chemistry of inclusions and matrix across the CO 3.0 - CO 3.7 metamorphic sequence.

*3.1.2 Brecciation and petrofabrics*

The CV3 and CO3 chondrites contain exotic matrix-rich lithic fragments called "dark inclusions" (e.g., Kracher et al., 1985; Johnson et al., 1990). We used the criteria of Weisberg and Prinz (1998) to distinguish lithic fragments from inclusions or clasts (chondrules, CAIs, AOAs). These criteria include dominance of fayalitic olivine (80 vol%) and textures distinct from surrounding matrix. Our measurements of inclusions and matrix in these meteorites include minimal contribution by lithic fragments, and we did not observe any chondrule- or aggregate-bearing dark inclusions. In the Leoville section, a prominent lithic fragment was masked off and not included in this study. At most, dark inclusions amount to less than a few volume % (Scott and Krot, 2003).

Many chondrites, particularly CV3, are known to exhibit petrofabrics. These range from weak but pervasive foliation in recrystallized matrix fayalite of Allende (Watt et al., 2006), to strong alignment of deformed chondrules in Leoville (Cain et al., 1986). Experiments indicate that hypervelocity impact shocks can cause both phenomena (Nakamura et al., 1995). Shock stages of the CV chondrites examined here vary slightly (Table 2; Scott et al. 1992). Rubin (2012) attributed differential porosity and alteration between reduced and oxidized CV chondrites to different previous histories of compaction and their later permeability to water. However, high porosity does not equate to high permeability (Bland et al., 2009). The relationship between porosity and shock stage in CV chondrites is not clear, nor is the timing of aqueous alteration relative to the timing of impact shocks. Here, we will assume that the CV chondrites Allende, Mokoia, Nova 002, and Vigarano have similar shock (i.e., compaction) histories (Table 2). We also assume that strong petrofabrics of chondrules (e.g., Leoville), which cannot be assessed correctly in single sections, have negligible effects on immediate measurements or the present conclusions drawn from those meteorites.

*3.1.3 Accretionary rims*

Chondrules and other inclusions are often surrounded by fine-grained accretionary rims (e.g., Rubin and Wasson, 1987; Huss et al., 2005). These rims may be distinguished from matrix in x-ray maps with varying degrees of difficulty. It has been proposed that some fraction, up to 100%, of the matrix in some chondrites was brought to parent bodies as fine-grained rims on clasts (Metzler et al., 1992). Brearley and Jones (1998, p.217) point out that rims in the least equilibrated CO chondrites represent material from the same reservoir as matrix. Here, because it is impossible to be consistent in distinguishing the boundaries between rims and matrix across large suites of inclusions and meteorites, we include fine-grained rims around inclusions as "matrix" (Huss et al., 2005).

**3.2. Sample Mapping**

Samples were analyzed using a Cameca SX100 electron probe microanalyzer (EPMA) equipped with a Bruker AXS Quantax 4010 energy dispersive spectrometer (EDS) and five wavelength dispersive spectrometers (WDS) at the AMNH. Four slabs and three polished sections of Allende, and 14 polished sections (Table 3) were mapped at a resolution of 13 microns/pixel (slabs) and 2-5 microns/pixel resolution (PS, PTS and selected areas in slabs). Slabs (6-8 mm thick) were constrained to fit in a 5.5 x 4 cm sample holder, and were pumped down in an $N_2$-purged bell jar for at least 24 hours prior to microprobe work. Intensities of K$\alpha$ x-ray emission for five elements (Si, Fe, Ca, Mg, Al or Mg, Ni, Ti, Al, Ca) were measured by WDS, and for three elements (S, Ni, Ti or S, Si, Fe) by EDS using Cameca software. Back-scattered electron (BSE) images were also collected. Beam mapping conditions are listed in Table 3 (for details see Electronic

Annex, Table S2). Only data for Si, Mg, Al, Ca, Ti, and Fe are presented or used in this study.

Mosaics of many contiguous x-ray map frames were collected on each sample surface by rastering stage motion in each $512^2$ pixel frame (Table 3, Fig. 1), keeping the beam nominally perpendicular to the surface. Time constraints limited collection to 8 elements in most cases, for short dwell times. Background counts were not collected. These conditions preclude full ZAF correction of the data. The spatial resolution of each map set was selected to optimize resolution and sample coverage within the EPMA time available. Map areas were set to overlap by at least one pixel on each edge.

**Table 3:** Sample data. Forms are unepoxied slabs, epoxied polished thick sections (PS), and epoxied polished thin sections (PTS). All samples are from the AMNH collection except Leoville (NMNH, USA) and Acfer 094 (Institut für Planetologie, Münster), as listed by sample number. Also listed are map resolution (res, μm/pixel), map conditions (accelerating voltage, keV; beam current, nA; dwell time, ms), total number of pixels (x $10^{-6}$), analyzed area A (mm$^2$), number of 512x512 pixel map frames, outline rim thickness (pixel) in outlined maps, and total number of items including isolated olivine and opaques (*n*) outlined in each map.

| Meteorite: | form | sample id. | res | kV | nA | ms | pxls x10$^{-6}$ | A(mm$^2$) | frames | rim | *n** |
|---|---|---|---|---|---|---|---|---|---|---|---|
| Allende | slab | 5046-1A | 13 | 15 | 40 | 10 | 10.70268 | 1808.75 | 34 | 1 | 1695 |
| Allende | slab | 5046-1C | 13 | 15 | 40 | 10 | 6.65954 | 1125.46 | 38 | 1 | 1093 |
| Allende | slab | 4884-s1B | 13 | 15 | 40 | 10 | 7.43929 | 1257.24 | 30 | 1 | 1013 |
| Allende | slab | 4884-s2B | 13 | 15 | 30 | 12 | 9.27800 | 1567.98 | 47 | 1 | 1397 |
| Allende | slab | 4884-s2B-F1 | 2 | 15 | 40 | 12 | 1.46093 | 5.84 | 6 | 0 | 175 |
| Allende | slab | 4884-s2B-F2 | 2 | 15 | 40 | 12 | 1.40949 | 5.64 | 6 | 0 | 118 |
| Allende | PS | AL2ps2 | 4 | 20 | 20 | 10+15 | 2.04442 | 32.71 | 9 | 0 | 228 |
| Allende | PS | AL2ps5 | 5 | 15 | 20 | 20 | 3.59244 | 89.81 | 23 | 0 | 466 |
| Allende | PS | AL2ps9 | 5 | 15 | 20 | 20 | 3.35842 | 83.96 | 24 | 0 | 371 |
| Allende | PTS | 4288-1 | 8 | 15 | 30 | 12 | 3.38267 | 216.49 | 15 | 0 | 372 |
| Tibooburra | PTS | 5003-1 | 6 | 15 | 30 | 12 | 8.04496 | 289.62 | 10 | 1 | 1300 |
| Nova 002 | PTS | 4826-2 | 7 | 12 | 30 | 12 | 1.91732 | 93.95 | 8 | 0 | 205 |
| Bali | PTS | 4936-1 | 12 | 15 | 40 | 12 | 1.34708 | 193.98 | 7 | 0 | 188 |
| Mokoia | PTS | 3906-4 | 9 | 15 | 40 | 12 | 1.03571 | 83.89 | 5 | 0 | 213 |
| Vigarano | PTS | 2226-4 | 7 | 15 | 30 | 15 | 2.01098 | 98.54 | 9 | 1 | 524 |
| Leoville | PTS | USNM-3535-1 | 10 | 15 | 30 | 15 | 1.48320 | 148.32 | 10 | 0 | 306 |
| | | | | | | total: | **65.167** | **7102.19** | | | |
| Meteorite: | form | sample id. | res | kV | nA | ms | pxls x10$^{-6}$ | A(mm$^2$) | frames | rim | *n** |
| Colony | PTS | 4595-1 | 2 | 15 | 30 | 20 | 8.12912 | 32.52 | 33 | 1 | 2578 |
| Kainsaz | PTS | 4717-1-1Cp1 | 2 | 15 | 30 | 20 | 5.90744 | 23.63 | 30 | 1 | 2491 |
| Kainsaz | PTS | 4717-1-1Cp2 | 2 | 15 | 30 | 20 | 5.60308 | 22.41 | 30 | 1 | 2633 |
| Ornans | PTS | 520-1-r4 | 1 | 15 | 20 | 20 | 14.01940 | 14.02 | 54 | 0 | 1044 |
| Lance | PTS | 618-1 | 3 | 15 | 30 | 15 | 5.09902 | 45.89 | 20 | 1 | 1948 |
| Warrenton | PTS | 4151-1 | 2 | 15 | 20 | 15 | 3.61709 | 14.47 | 14 | 0 | 662 |
| Acfer 094 | PTS | IfP-PL93022 | 1 | 15 | 20 | 15 | 10.45333 | 10.45 | 41 | 1 | 1134 |
| | | | | | | total: | **52.828** | **163.39** | | | |

* The total number of outlined items including isolated olivine and metallic grains.

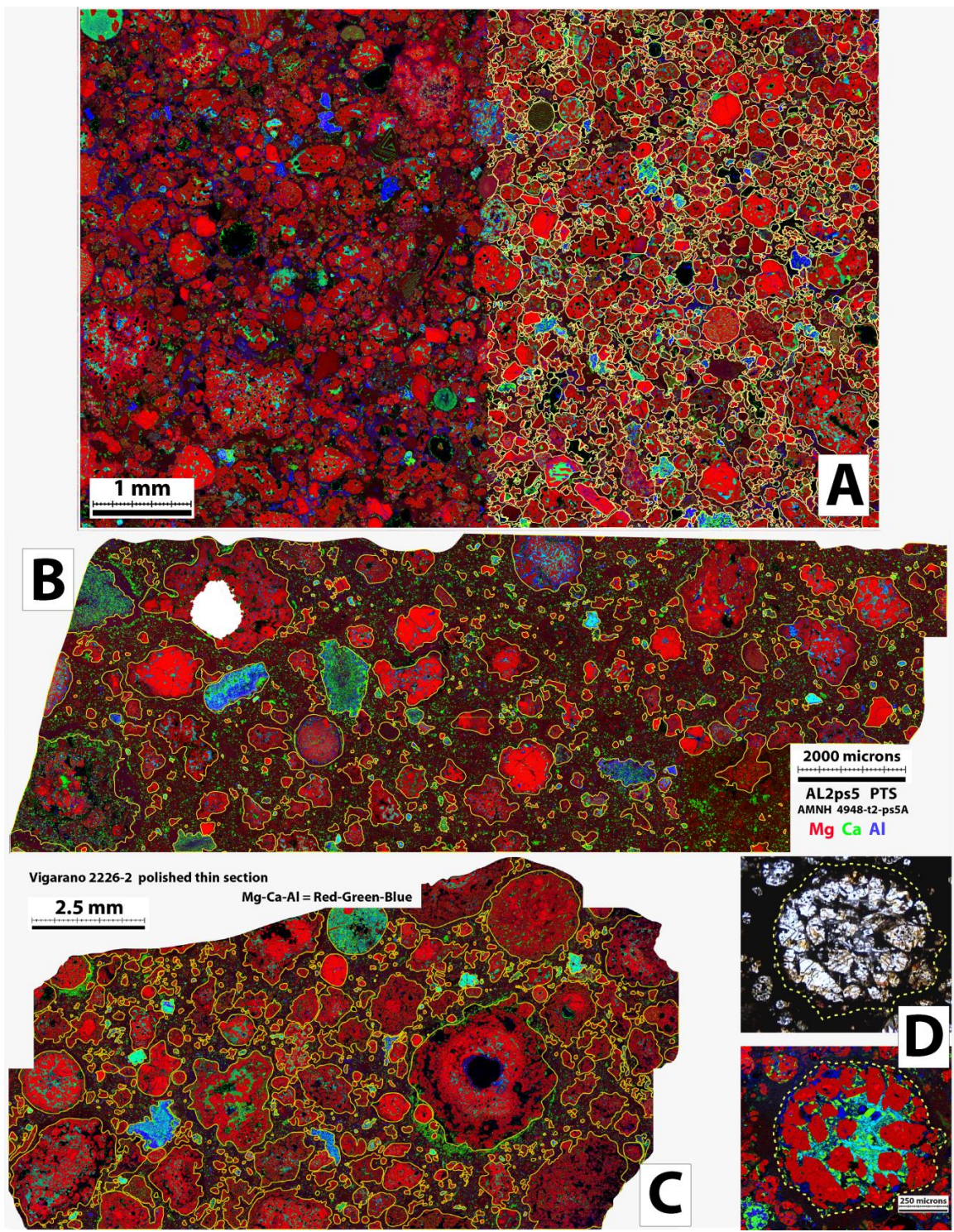

**Fig. 1:** Mg-Ca-Al = red-green-blue x-ray mosaics of (A) Kainsaz 4717-1Cp1 (CO 3.2), showing clast outlines (yellow lines) in right hand portion; (B) Allende AL2ps25 (CV-OxA ~3.6), showing outlines and clasts labels; (C) Vigarano (CV-R 3.3), showing outlines; (D) ~800 µm chondrule in Colony (CO 3.0), plane polarized light image (upper) and x-ray mosaic (lower) showing outline (dashed line);

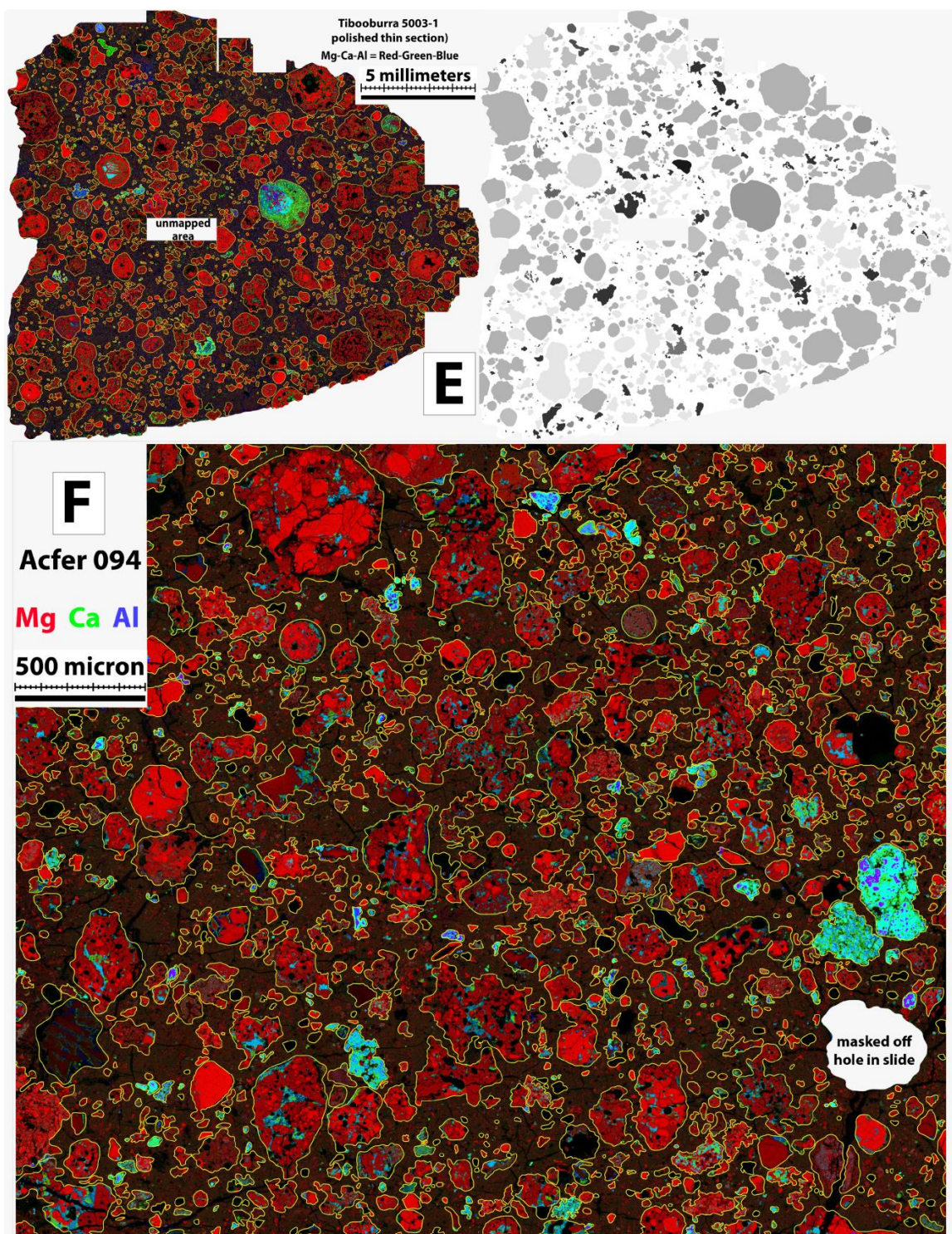

**Fig. 1, continued:** (E) Tibooburra (CV 3), showing outlines and, on right, grayscale input to ImageJ step of analysis algorithm; and (F) cropped area of Acfer 094 (C2-ungr) showing outlines (yellow).

The beam spot size was always set to the minimum (~1 µm) diameter, however excitation volume was likely greater than 1 µm$^3$. Our goal was to be able to "mine" each pixel of data for its chemistry and mineralogy. Previous investigations have analyzed matrix for the elements mapped in this study using large beam spots (e.g., 100 µm, McSween and Richardson, 1977; 30 µm, Hezel and Palme, 2010) or rasters (e.g., 30x30µm, Grossman and Brearley, 2005). If we have correctly segmented images to isolate inclusions from matrix, then our matrix analysis technique is identical to rastering over a very large area by stage motion, rather than by beam deflection. Zolensky et al. (1993) used a focused beam, and noted that their results compared well with those reported by McSween and Richardson (1977). As pointed out by Grossman and Brearley (2005), matrix grain size is likely small relative to the excitation volume of the electron beam, reducing the need for density corrections.

For Allende slabs mapped at 13 µm/pxl resolution, the flatness of the surface varied sufficiently over sample areas that take-off angle variations systematically affected x-ray intensity measurements from edge to edge of each map frame. Automatic corner focusing was not used, to avoid software failure during long automated sessions. Each map in each mosaic was focused in reflected light at the center of the map at the highest optical resolution. Centers of each map frame (in µm) were recorded for later, off-line stitching of frames. Map sets stripped of scale bars, labels and look-up table information were output in 32-bit tagged image file (tif) format to preserve original intensity information. Only Mokoia (3906-4) was restricted to 8-bit output.

X-ray intensity maps for each element were stitched together to form a seamless single 32-bit grayscale mosaic. For reference, 24-bit red-green-blue (RGB) composite maps in three-element combinations (e.g., Mg-Ca-Al, Si-Ca-Fe) were also produced (Fig. 1). These RGB composite images were color-balanced using a method identical to that for automatic color leveling used in Adobe Photoshop™ (v. 9). All of these steps were done using a single custom-written IDL (IDL, 2015) program, retaining original (100 dpi) image resolution, non-interlaced, throughout.

We present most data as counts per pixel. Our best calibration point is the bulk compositions of the chondrites (Table 1). Our x-ray intensity maps were calibrated post-collection against similar data collected on standards. We have established in related work (Crapster-Pregont and Ebel, 2014a) that changes in microprobe instrument characteristics over time have only slight effects on such calibrations. For purposes of comparison, x-ray intensity counts are closest to the original data. In this work, we limit presentation to these data.

**3.3. Image Analysis**

*3.3.1. Segmentation of chondrules and refractory inclusions*

All the element maps, and several RGB composites, were imported as registered (exactly overlaid) layers into drawing software (Adobe Illustrator™, AI). The parts of images that did not contain sample, or were otherwise compromised, were masked off with a separate layer of a specific grayscale, to prevent consideration of these regions in subsequent image analysis (Fig. 1b, 1c). Registered (stacked) maps are analogous to a remotely-sensed hypercube or multi-spectral image of a planetary surface. They allow rapid cycling between element and composite maps of the sample surface, facilitate hand-drawing of outlines around each of the clasts present in each surface, and inform the identification of each clast by type. This step, segmentation, was performed using drawing tools in the AI software. Segmentation was performed largely by student coauthors, with outlines and identification edited extensively by senior authors. Inclusion categories and identification criteria are listed in Table 4. We did not distinguish between Type I (FeO-poor) and Type II (FeO-rich) chondrule types (Patzer et al., 2012), although the data could be mined for this information. All outlines were adjusted to fit inside the mask layer (e.g., Fig. 1a). Figure 1 illustrates selected Mg-Ca-Al = red-green-blue mosaics, with outlines of inclusions. Figure 1d illustrates the transmitted light and x-ray mosaic appearance of a single outlined porphyritic olivine chondrule from Colony (CO 3.0).

**Table 4:** Categories of inclusions and other objects.

| **Chondrules** | | Round to sub-round. Dominated by olivine and pyroxene. |
|---|---|---|
| | **BO** | Barred Olivine chondrules. |
| | **RP** | Radiating Pyroxene chondrules. |
| | **PO** | Porphyritic Olivine chondrules ($\geq 90$ olivine). |
| | **POP** | Porphyritic Olivine - Pyroxene chondrules. |
| | **PP** | Porphyritic Pyroxene chondrules ($\geq 90\%$ pyroxene). |
| | **MCH** | Miscellaneous Mg-silicate-rich inclusions. |
| **Refractory Inclusions** | | Dominated by Ca-, Al-rich phases. Olivine is minor to absent. |
| | **CCH** | Al-rich chondrules "Type C". Minor olivine. |
| | **A-CAI** | Compact and Fluffy Type A CAI are counted as one class. CTA contain no matrix, inside a continuous, well defined border. FTA contain matrix in embayments or islands. |
| | **B-CAI** | Type B CAI. Round to sub-round, igneous texture, often rimmed. |
| | **MCAI** | Miscellaneous Ca-, Al-rich inclusions . |
| | **AOAs** | Amoeboid Olivine Aggregates. Irregular shape, nodular structure, CAI-like nodules rimmed by olivine. |
| **Other** | **IOL** | Isolated olivine grains or aggregates in matrix. |
| | **FENI** | Isolated metal or metal sulfide grains in matrix. |
| | **DI** | Dark Inclusions. Fine grained exotic lithic fragments. |

In each AI layer containing outlines of all instances of a specific inclusion type, the object fill color was set to a unique grayscale value indicative of that inclusion class. Because AI allows drawing at sub-pixel resolution, one-pixel wide borders on each object outline were set to the same grayscale (white) reserved for meteorite matrix for processing selected sample maps ('rim', Table 3). This was done so that two or more outlined objects would not appear to overlap, and thus be segmented as a single object in later steps. A non-anti-aliased 8-bit tif file was then exported, containing the mask and the complete set of filled clast layers, with each object type (e.g., BO, AOA, etc.) filled with a separate grayscale pixel value (e.g., Fig. 1e).

Second, the masked grayscale object file was imported into the open source ImageJ (2013) software. After setting image properties (μm/pixel), the masked grayscale image was thresholded so that all the outlined objects (clasts), but not mask or matrix, were selected. This thresholded image was analyzed for particle size, area, center of mass, and other parameters, and a drawing was output with the outline of each clast, each with a unique identifying number referencing its properties in an output table. In cases with one-pixel rims (Table 3), apparent objects containing five or fewer pixels were discarded as artifacts of the method. Where center of mass fell outside the object boundaries, the center of mass coordinates were corrected manually to be within the object.

*3.3.2. Analysis of segmented images*

At this stage, the suite of single-element grayscale EPMA x-ray mosaics, the mask layer, the masked grayscale object map (e.g., Fig. 1e), and the table of image data were all imported into another software module written in IDL. First, a histogram was produced, and the number of pixels of each grayscale corresponding to a particular object type was counted. This yielded the relative proportions of each object type and matrix, as well as the total area measured (Tables 5, 6, 7).

Next, the pixels belonging to each individual outlined clast or inclusion were identified. An initial pass allowed thresholding each object with a unique combination of red-green-blue (RGB) color, using a region-growing algorithm. In this algorithm, the center of mass pixel for each object was grown by addition of neighbors, until the edge of the object was reached, whereupon addition of new pixels ceased. Objects for which the center of mass was in matrix were flagged in this step by software and corrected manually by alteration of the center of mass data table created in step two. This step yielded the number of pixels in each clast, with each clast identified by type. This step also yielded a map of the same size as the original element maps, with all pixels belonging to each inclusion having a unique RGB color, and a table with, for each inclusion, the x-y coordinates of the center of mass, RGB color, and x-y (max-min) coordinates of a rectangle containing all the pixels in that inclusion. With this information, each separate inclusion could be queried efficiently on a pixel-by-pixel basis.

Fourth, each pixel in each inclusion (and also matrix) was analyzed for the relative abundances of elements measured in the x-ray maps. Pixels with a low sum of major element x-ray intensities were labeled as holes (part of the mask), and were not considered further. The total intensity of emission for each element over the area of each inclusion, minus holes, was computed separately for each inclusion and for matrix. The total intensities of major elements per pixel in each object should approximate the bulk chemistry of that object. The total intensity across all pixels in a sample should directly approximate the bulk composition of the sample. These results are summarized in Table 8.

Finally, each pixel was queried for mineralogy, resulting in a modal analysis. This proved to be the least reliable step, given the short dwell times used to collect the x-ray map data. The lack of epoxy, coarse resolution, and slight tilt of the large Allende slabs rendered their element data unreliable for quantitative treatment, so analysis of integrated element intensities was abandoned for Allende slabs. All this data is available in table form, e.g., the integrated major element abundances over all pixels in each of 2471 inclusions, and matrix, outlined in Colony (AMNH 4595-1).

*3.3.3. Analysis of matrix heterogeneity*

Matrix heterogeneity was assessed by virtually dividing maps into arbitrarily narrow slices and measuring the mean count/pixel of each element in the matrix portion of each slice. Heterogeneity was measured in samples Allende AL2ps5 and AL2ps9, Colony 4595-1, Kainsaz 4717-1-Cp2 and Cp2, and Acfer 094, each divided into 3, 6, 9, and 27 vertical slices (Fig. S1).

**3.4. Multi-scale Modal Analysis**

The scale required to map the four large slabs of Allende was relatively coarse (13 μm/pxl). Small objects within the matrix of Allende are unidentifiable at this scale. To correct for this, two areas in slab 4884-s2B that had already been analyzed at 13 μm/pxl were chosen for a second analysis at 2 μm/pxl resolution. These areas (F1, F2 in Table 3) were subject to the same protocol as all other samples. Large objects previously counted in these particular areas were disregarded, and the modal abundances of small inclusions were counted. These results were used to correct the matrix and inclusion modal abundances obtained by coarse analysis of the large slabs. Data for inclusion size was not corrected in this way, so there is a systematic difference between our size result for Allende slabs and polished sections.

**3.5. Size Analysis**

While 2D surface abundance is directly proportional to the relative volumetric abundances of clasts (Delesse, 1848; Chayes, 1956), the 2D analysis must be corrected for biases to yield accurate size distributions of clasts (Dodd, 1976; Martin and Mills, 1976). Previous workers (e.g., Rubin, 1989) have used the size scale of sedimentology (Folk, 1980), where $\phi = -\log_2$ (diameter, mm). A random slice through an aggregate of rimmed, spherical clasts in matrix (e.g., a thin section) yields apparent diameters smaller than true diameters, and rims that are thicker than true rims. However, random slices also over-sample large clasts. The accuracy of 2D-3D corrections is the subject of active research (e.g., Paque and Cuzzi, 1997; Friedrich, 2013). We report primarily relative abundance size histograms (as in Friedrich et al. 2014), but we have also applied the corrections described by Eisenhour (1996) for 1) non-equatorial sectioning of clasts, and 2) the unequal probability of sectioning different size clasts in large aggregates. The first correction uses a finite difference method of back substitution (after Saltykov, 1967; Hughes, 1978), here using a log-constant bin width of $\phi/4$, and directly applying equation 3 of Eisenhour (1996). We follow this equation to $i = 30$ for each bin. The second correction is based on the fact that the probability of intersecting a clast is proportional to its size, so the relative abundances of clasts with diameters $d_a$ and $d_b$, and measured abundances $N_a$ and $N_b$, is $N_a d_b / N_b d_a$. We verified our application of these algorithms by recomputing the data of Rubin (1989), following Eisenhour (1996, his Table 1).

## 4. RESULTS

### 4.1. Inclusion and Matrix Abundance

Modal abundances of components in CV chondrites are presented in Table 5, and for CO chondrites in Table 6, with a comparison to earlier work (McSween, 1977a, b, c). We also present the first such results for Acfer 094, an ungrouped chondrite with CO affinities. We do not compare abundances with King and King (1978) because they did not count inclusions of < 0.1mm apparent diameter. Ratios of matrix (including isolated olivine, metal, and dark inclusions) to inclusions (chondrules, refractory inclusions) are calculated. The wide diversity in matrix/inclusion ratios in our data is illustrated in Figure 2. Nova 002 is slightly weathered, and has been described as an oxidized subtype (Treiman and DeHart, 1992). Our data indicate Nova 002 has strong affinity with the reduced CV. Allende polished sections (PS, PTS) yielded results consistent with analyses of large slabs, suggesting that the relative abundances of components in the other CV chondrites is a robust measurement, despite the limit to single sections. The Allende result also suggests that the modal abundance measurement is not affected by our inability to segment small inclusions in the slabs imaged at 13 μm/pxl.

Our findings for CV are broadly consistent with those of McSween (1977b), except that matrix is slightly more abundant in our result, with a correspondingly lower abundance of chondrules and CAIs (Table 5). This difference is likely due to different objective criteria used in the studies, and biases intrinsic to x-ray map analysis relative to optical point counting techniques (Fig.1d). In x-ray and BSE maps the accretionary rims on many chondrules appear similar to matrix, so rims were preferentially included here as matrix. For example, our matrix abundance (29.6%, Table 5) for Leoville (CV-R) falls between that of McSween (1977b, 35.1% matrix), who used optical methods, and Patzer et al. (2012, 23% matrix) who used x-ray element maps.

### 4.2. Diversity of inclusion compositions and homogeneity of matrix

Chondrules and refractory inclusions showed considerable scatter among individual inclusions in all comparisons of major element distributions (e.g., Fig. 3; Electronic Annex**,** Figs S3 - S12). There is an enormous variety in element abundance among chondrules, CAIs, and AOAs, even those with similar textures. Nevertheless, the mean compositions of various types of inclusions follow predictable patterns. There is a clear distinction between pyroxene-rich (PP), mixed (POP), and olivine-rich (PO) chondrules (e.g., Fig. 3). Their average value ("chondrules") is complementary to matrix, indicating that the chondrules are themselves complementary in combining (over a huge range of compositions) just so as to make, with matrix, a chondrite.

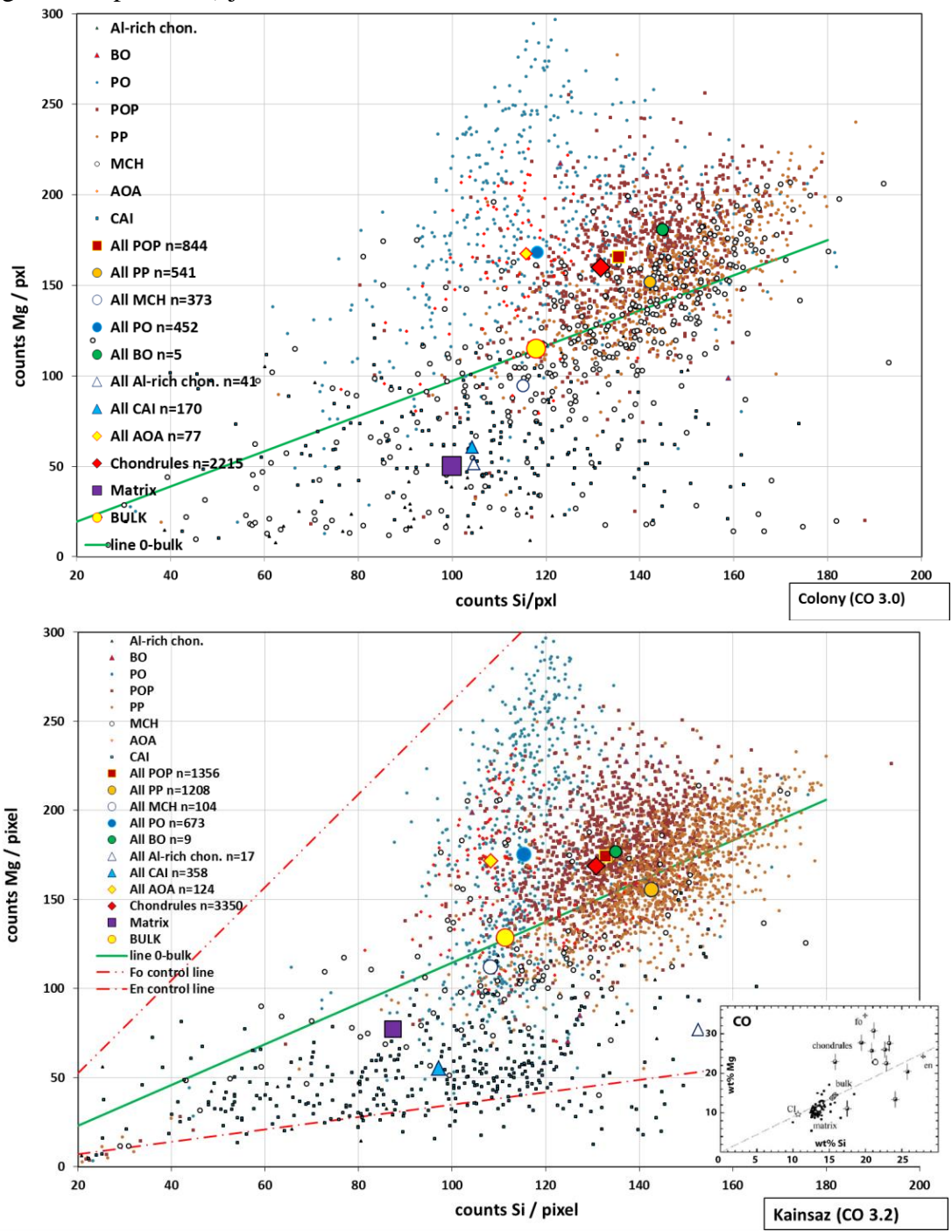

**Fig. 3:** Per pixel counts of Si and Mg for all clasts, mean clasts, matrix, and bulk meteorite for Colony (CO 3.0) and Kainsaz (CO 3.2). Inset is from Hezel and Palme (2010, their Fig. 1). Lower plot shows Mg/Si count ratios from $128^2$ pixel maps of enstatite (En, $MgSiO_3$, red dash-dot) and forsterite (Fo, $Mg_2SiO_4$, red dash-dot-dot) standards, normalized to conditions of Kainsaz analysis.

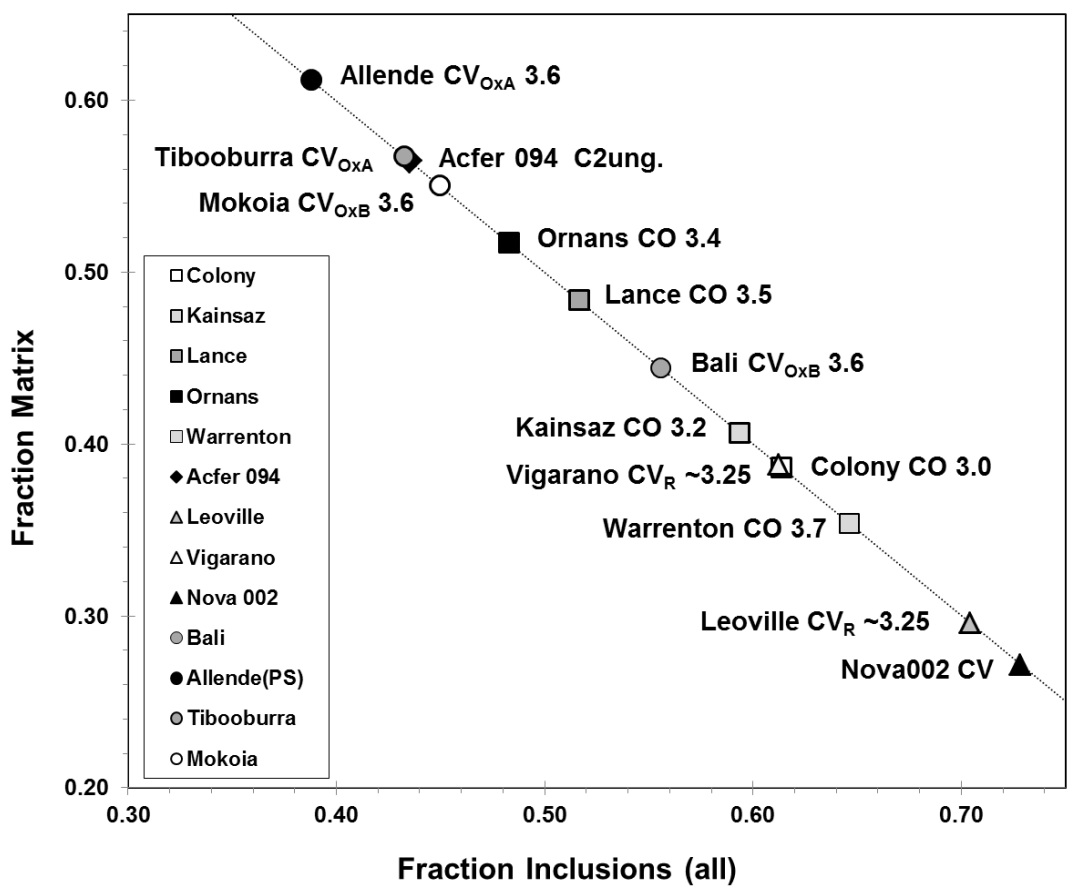

**Fig. 2:** Fraction of pixels (area) in matrix and inclusions (chondrules, CAIs, other clasts). All sum to unity (line). The wide spread of clast/matrix ratios is apparent.

Matrix heterogeneity about the mean increased slightly with the number of slices taken (3, 6, 9, and 27). In Colony and Kainsaz, which had been polished with alumina abrasive, Al shows the greatest heterogeneity, with ~12% variability across a set of 6, 9, or 27 slices. However, Al also shows the highest heterogeneity in the Allende section AL2ps5, which was polished with diamond paste. In general, variations among element count/pxl values for each slice increased only slightly with the number of slices taken (Fig. S2). Matrix appears to be chemically homogeneous in the elements analyzed in our samples.

### 4.3. Refractory Inclusion Abundance

Our results provide a robust count of the relative abundances of AOAs and CAIs in Allende. CAIs are slightly more abundant than AOAs in Allende, and CAIs are <4.8%, consistent with the 2.98 area% of Hezel et al. (2008). The modal abundance of CAIs in Acfer 094 (2.53%, Table 6) is slightly higher than the < 2% reported by Weber (1995). Our findings underscore the importance of AOAs to the origin of meteoritic chondrules (Grossman and Steele, 1976; Weisberg et al., 2004; Krot et al., 2004b).

**Table 5:** Relative abundance of inclusions in CV chondrites, combined for comparison with those reported by McSween (1977b). Pixel percentages of major components are given, then the number (n) of items of all types, the number ratio of chondrules to CAIs, and the (*) matrix/inclusion pixel ratio (matrix + IOL + ms)/(CAI + AOA + chondrules). McSween (1977b, his table 2) ratios are (**) calculated as matrix/(lithic fragments + CAI + AOA + chondrules). Nova002 data is from Treiman and DeHart (1992). The 'std dev' is 1σ. Area-weighted mean is an average weighted by µm² area measured (Table S1), dominated by Allende.

| This work: | Allende | Tibooburra | Nova 002 | Bali | Mokoia | Vigarano | Leoville | mean | std dev | A-wtd mean |
|---|---|---|---|---|---|---|---|---|---|---|
| matrix+lithic frags. | 55.24 | 56.40 | 35.85 | 43.33 | 54.11 | 38.65 | 29.19 | 44.68 | 10.75 | 53.92 |
| IOL | 1.37 | 0.23 | 0.05 | 0.33 | 0.18 | 0.06 | 0.21 | 0.35 | 0.46 | 1.22 |
| CAI | 4.32 | 0.83 | 0.83 | 5.70 | 4.90 | 3.11 | 1.62 | 3.04 | 2.00 | 4.10 |
| AOA | 3.48 | 2.11 | 3.73 | 1.87 | 2.00 | 1.47 | 4.03 | 2.67 | 1.04 | 3.35 |
| chondrule | 35.46 | 40.32 | 59.07 | 48.13 | 38.02 | 56.66 | 64.76 | 48.92 | 11.46 | 37.26 |
| metal + sulfide (ms) | 0.12 | 0.11 | 0.47 | 0.64 | 0.78 | 0.06 | 0.19 | 0.34 | 0.29 | 0.15 |
| n inclusions | 5516 | 1233 | 181 | 167 | 179 | 507 | 267 | | | |
| matrix/inclusion* | 1.31 | 1.31 | 0.57 | 0.80 | 1.23 | 0.63 | 0.42 | 0.90 | 0.38 | 1.26 |
| n(chondrule)/n(CAI) | 4.60 | 6.45 | 8.05 | 3.00 | 4.38 | 5.34 | 3.00 | 5.0 | 1.83 | 4.65 |
| **McSween (1977b)** | Allende | | Nova 002*** | Bali | Mokoia | Vigarano | Leoville | mean | std dev | |
| matrix | 38.4 | | 33.1 | 50 | 39.8 | 34.5 | 35.1 | 38.5 | 6.18 | |
| Lithic & Mineral Frags. | 2.9 | | 0.8 | 1 | 3.6 | 2.1 | 1.1 | 1.9 | 1.15 | |
| CAI | 9.4 | | 8.3 | 4 | 3.5 | 5.3 | 6.6 | 6.2 | 2.36 | |
| AOA | 3.2 | | 0.6 | 7.5 | 2.7 | 5.4 | 1.3 | 3.5 | 2.59 | |
| chondrule | 43 | | 56.4 | 32 | 46.9 | 48.9 | 51.4 | 46.4 | 8.37 | |
| opaque phases | 3.1 | | n.d. | 5.5 | 3.5 | 3.0 | 4.5 | 3.9 | 1.06 | |
| matrix/inclusion** | 0.66 | | 0.50 | 1.12 | 0.70 | 0.56 | 0.58 | 0.7 | 0.23 | |
| n points | 1572 | | n.a. | 1675 | 1510 | 1594 | 1705 | | | |

* 'chondrules' here includes CCH (Al-rich); matrix/inclusion = (matrix+IOL+ms)/(CAI+AOA+chondrules)
** McSween (1977b) calculates this ratio as: matrix/(lithics+CAI+AOA+chondrule)
*** Treiman and DeHart (1992)

Hezel et al. (2008) presented strong arguments regarding the representative sampling of CAIs. The sizes of CAIs in the CO chondrites studied here are small, below the average radius of 113 µm and maximum 500 µm considered in the model of Hezel et al. (2008). Assuming a Poisson distribution with 2.61 area% CAIs (their Fig. 2), Hezel et al. determined that only 23% of 25 mm², and 40% of 100 mm² square areas would contain that 2.61% modal abundance, with more samples biased to high area% than to low. They showed that an area of 2500 mm² would be required for any sampling to be representative. However, the mean apparent radius of all clasts in CO chondrites is ~42 µm, with 1σ of 36 µm ($n = 9043$), with CAIs 29 µm (1σ = 23, $n = 846$). Our CO results show no correlation between area measured and CAI area%. From their fitting of a Poisson distribution to their own and existing data, Hezel et al. (2008) arrived at a CAI abundance of 0.99 area% for CO chondrites, and 1.12 for Acfer 094. Our result is 2.4 ± 0.29 area% for CO, and 2.53 for Acfer 094. Our result for Leoville (CV3) CAI abundance, 1.62% in 148 mm², is unlikely to be representative. It is below those of Patzer et al. (2012), who found 3.1% in 220 mm², and McSween (1977b) who reports 6.6%.

CAI abundances in slabs and sections differ, and this may depend on measurement resolution. Typical mean inclusion radii are 600 µm in Allende slabs analyzed at 13 µm/pxl resolution, and our results include 5760 mm$^2$ of slab area. Apparent mean inclusion radii are significantly smaller in the CV we measured in thin sections, due to the much higher spatial resolution (Table 3). Lack of map resolution, despite multi-scale correction, is a possible reason for the difference between the overall Allende CAI abundance (4.32%) and the mean CV value of 2.63 ± 1.62. The Allende slabs contain 4.31% CAIs and 3.5% AOAs, while the Allende sections (423 mm$^2$) contain 4.49% CAIs, and 3.15% AOAs.

**Table 6:** Relative abundances of inclusions in CO chondrites, combined for comparison with those reported by McSween (1977a). Notes are as for Table 5. Mean and area-weighted mean exclude Acfer 094.

| This work: | Colony | Kainsaz-all | Ornans | Lancé | Warrenton | mean | std dev | A-wtd mean | Acfer 094 |
|---|---|---|---|---|---|---|---|---|---|
| matrix+lithic frags. | 37.42 | 37.13 | 47.07 | 45.78 | 33.09 | 40.1 | 6.04 | 41.17 | 54.54 |
| IOL | 0.38 | 0.38 | 2.55 | 0.60 | 0.86 | 1.0 | 0.91 | 1.16 | 0.55 |
| CAI | 2.19 | 2.52 | 2.64 | 2.33 | 2.06 | 2.3 | 0.23 | 2.43 | 2.53 |
| AOA | 4.40 | 4.38 | 2.35 | 2.24 | 4.39 | 3.5 | 1.15 | 3.45 | 3.78 |
| chondrules* | 55.32 | 52.44 | 43.35 | 47.00 | 58.15 | 51.3 | 6.05 | 49.82 | 37.19 |
| metal + sulfide (ms) | 0.29 | 3.15 | 2.05 | 2.06 | 1.45 | 1.8 | 1.05 | 1.96 | 1.40 |
| *n* inclusions | 2502 | 3849 | 610 | 1569 | 513 | | | | 833 |
| matrix/inclusion* | 0.61 | 0.63 | 1.03 | 0.90 | 0.53 | 0.7 | 0.21 | 0.78 | 1.27 |
| *n*(chondrule)/*n*(CAI) | 13.26 | 9.41 | 4.96 | 7.92 | 8.21 | 8.8 | 3.01 | 8.40 | 3.94 |
| **McSween (1977a)** | Colony** | Kainsaz | Felix (3.3) | Ornans | Lancé | Warrenton | mean | std dev | Acfer 094*** |
| matrix | 29.3 | 30 | 32.9 | 40.1 | 32.8 | 37.8 | 33.8 | 4.29 | |
| Lithic & Mineral Frags. | 23.4 | 7.6 | 9.3 | 7.8 | 7.4 | 6.8 | 10.4 | 6.43 | |
| CAI | 3.3 | 2.4 | 3.5 | 1.4 | 1.2 | 2.5 | 2.4 | 0.95 | |
| AOA | 7.85 | 7.4 | 8.4 | 11 | 11.8 | 11.6 | 9.7 | 2.01 | |
| chondrule | 32.95 | 45.2 | 39.7 | 34.3 | 40.9 | 38.1 | 38.5 | 4.49 | |
| opaque phases | 3.2 | 7.4 | 6.2 | 5.4 | 5.9 | 3.5 | 5.3 | 1.63 | 6.4 |
| matrix/inclusion** | *0.43* | 0.43 | 0.49 | 0.67 | 0.49 | 0.61 | 0.52 | 0.10 | |
| *n* points | ~2000 | 1765 | 1529 | 2711 | 1533 | 2852 | | | |
| vol% matrix*** | 28.8 | 24.8 | 32.8 | n.d. | 26.9 | n.d. | | | 42 |
| matrix/inclusion*** | 0.40 | 0.33 | 0.49 | | 0.37 | | | | 0.72 |

\* 'chondrules' here includes CCH (Al-rich); matrix/inclusion = (matrix+IOL+ms)/(CAI+AOA+chondrules)

\*\* Reported (H.Y. McSween, pers. comm.) by Rubin et al. (1985, their Table 1, mean columns 2+3). All other data from McSween (1977a), calculated as: matrix/(lithics+CAI+AOA+chondrule)

\*\*\* Grossman and Brearley (2005) Table 8. Chondrules + CAI + AOA = 56.1 vol% calculated by difference in Acfer 094.

### 4.4. Chondrule Type Distribution

Each chondrule in each sample was assigned a type, based on visual texture and estimated mineralogy. Table 7 presents relative abundances as % of all chondrules and CAIs, where inclusion type was determined by inspection. An alternative method of distinguishing between porphyritic chondrule types (PO, POP, and PP) could include machine determination of the olivine/pyroxene ratio in each chondrule. This could be accomplished using our data sets, however, results (Fig. 3) illustrate that visual estimates of type generally place these chondrules in the appropriate Mg-Si composition region.

### 4.5. Compositions of Components

21The fundamental unit for the data in this paper is element counts per pixel. This unit can in principal be linearly mapped to element mass fraction. Because the large area maps were collected at opportune times of instrument availability, and before we were entirely aware of the complexity of certain mapping issues, difficulties remain in directly correlating counts/pixel to mass%. Temporal changes to collection conditions include spectrometer window contamination, filament heat, spectrometer drift, and other long-term variations in EPMA characteristics. For samples with large pixel counts (e.g., $10^6$ to $10^9$), we assume that bulk x-ray intensities (count/pxl) on an entire sample correlate directly to bulk composition by any other measure (e.g., wt%) if backgrounds are small. Although data on various CO and CV chondrites were collected with differing conditions (Electronic Annex, Table S2), we assume that we may normalize each chondrite bulk composition to a particular sample (Colony for CO, Allende for CV). We may then use these normalization factors to correct matrix and clast values across meteorite suites. The underlying assumptions are that all of the CO chondrites (3.0 to 3.7) have essentially identical bulk compositions, and that the CV chondrites do as well. Published data support this assumption (e.g., Hezel and Palme, 2010; above, section "Bulk Compositions"; Table 1).

**Table 7:** Relative abundances of chondrule and CAI types in CV and CO chondrites. Radiating pyroxene and cryptocrystalline abundances are combined in comparing our results to Rubin (1989, his Table 2; cf. Eisenhour 1996, his Table 1).

|  | Allende | Tibooburra | Nova 002 | Bali | Mokoia | Vigarano | Leoville | mean | stdev |
|---|---|---|---|---|---|---|---|---|---|
| Barred olivine | 3.83 | 1.64 | 3.86 | 0.76 | 0.94 | 0.67 | 0.09 | 1.69 | 1.54 |
| Radiating pyroxene | 0.13 | 0.00 | 0.00 | 0.00 | 0.00 | 0.00 | 0.00 | 0.02 | 0.05 |
| Porphyritic olivine | 40.91 | 8.49 | 9.21 | 27.07 | 29.09 | 7.94 | 29.11 | 21.69 | 13.08 |
| Porphyritic olivine-pyroxene | 36.38 | 62.51 | 63.12 | 64.88 | 48.95 | 84.06 | 55.65 | 59.37 | 14.80 |
| Porphyritic pyroxene | 1.40 | 1.20 | 7.35 | 2.30 | 0.40 | 4.05 | 1.11 | 2.55 | 2.42 |
| Al-rich | 1.40 | 3.83 | 3.71 | 1.12 | 0.00 | 0.00 | 0.85 | 1.56 | 1.60 |
| unclassified | 15.95 | 22.33 | 12.75 | 3.87 | 20.62 | 3.27 | 13.18 | 13.14 | 7.44 |
| Type B CAI/total CAI | 8.93 | 5.43 | 7.66 | 0.00 | 0.00 | 0.95 | 3.07 | 3.72 | 3.68 |

|  | Colony | Kainsaz | Ornans | Lance | Warrenton | mean | stdev | Rubin 1989* | Acfer 094 |
|---|---|---|---|---|---|---|---|---|---|
| Barred olivine | 0.81 | 1.49 | 4.15 | 0.55 | 0.06 | 1.41 | 1.62 | 2 | 1.81 |
| Radiating pyroxene | 0.20 | 0.00 | 0.00 | 0.00 | 0.00 | 0.04 | 0.09 | 3 | 0.00 |
| Porphyritic olivine | 18.83 | 14.14 | 14.63 | 16.62 | 26.55 | 18.15 | 5.04 | 8 | 10.63 |
| Porphyritic olivine-pyroxene | 60.36 | 63.01 | 73.43 | 59.51 | 64.96 | 64.26 | 5.56 | 69 | 47.71 |
| Porphyritic pyroxene | 11.96 | 16.08 | 3.37 | 19.48 | 2.72 | 10.72 | 7.50 | 18 | 24.65 |
| Al-rich | 1.30 | 0.80 | 3.33 | 0.26 | 0.24 | 1.19 | 1.28 | - | 0.00 |
| unclassified | 6.55 | 4.48 | 1.08 | 3.57 | 5.46 | 4.23 | 2.08 |  | 15.21 |
| Type B CAI/total CAI | 0.00 | 2.29 | 29.67 | 3.32 | 3.75 | 7.81 | 12.31 |  |  |

For the first time, we are able to quantify the total fraction of each major element accounted for by refractory inclusions, matrix and chondrules (Table 8). The methods we employ allow direct integration of the x-ray intensities of each pixel in each inclusion (clast), and in matrix. The sum over all pixels in an inclusion, over the total number of pixels in that inclusion, or all matrix pixels, or over the entire section, may be compared. Table 8 presents such data as the fractional number of total counts in each chondrite accounted for by refractory inclusions separately (AOA, CAI, Al-rich chondrules), ferromagnesian chondrules as a group (BO, PO, POP, PP, and unclassified chondrules MCH), and matrix (including isolated olivine grains, dark inclusions, and isolated metal grains). Data for Allende (PS) are a combination of polished sections AL2ps5 and AL2ps9 only, since those data were collected under identical conditions.

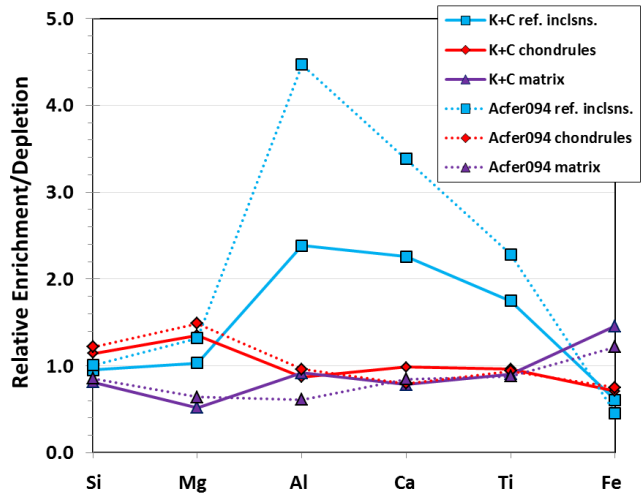

**Fig. 4:** Relative enrichment/depletion of elements among components in CO chondrites and Acfer 094. The fraction of total (bulk) element counts accounted for by each clast type is divided by the total number of pixels in each clast type. Shown are combined data for Colony and Kainsaz (K+C, solid lines) compared to data for Acfer 094 (dashed). "Ref. inclsns." includes CAI + AOA + Al-rich chondrules, the most Ca-, Al-rich inclusions.

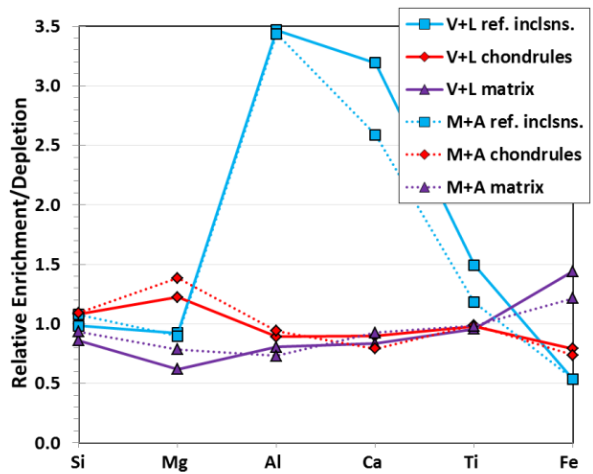

**Fig. 5:** Relative enrichment/depletion of elements among components in CV chondrites. The fraction of total (bulk) element counts accounted for by each clast type is divided by the total number of pixels in each clast type. Shown are combined data for reduced CV Vigarano and Leoville (V+L, solid lines) and oxidized CV Mokoia and Allende (M+A, dotted). Notes as for Fig. 4.

**Table 8:** Element distribution among components in CV and CO chondrites, and Acfer 094 (C2-ungr).

|  | Si | Mg | Al | Ca | Ti | Fe | area |
|---|---|---|---|---|---|---|---|
| Allende(PS) CAIs | 0.038 | 0.021 | 0.111 | 0.117 | 0.057 | 0.021 | 0.036 |
| Allende(PS) AOAs | 0.021 | 0.024 | 0.023 | 0.021 | 0.022 | 0.020 | 0.021 |
| Allende(PS) Al-rich chon. | 0.025 | 0.019 | 0.060 | 0.041 | 0.030 | 0.007 | 0.019 |
| Allende(PS) chondrules | 0.351 | 0.455 | 0.292 | 0.242 | 0.310 | 0.220 | 0.313 |
| Allende(PS) matrix | 0.565 | 0.480 | 0.513 | 0.579 | 0.581 | 0.732 | 0.612 |
|  |  |  |  |  |  |  |  |
| Tibooburra CAIs | 0.007 | 0.004 | 0.044 | 0.039 | 0.009 | 0.003 | 0.008 |
| Tibooburra AOAs | 0.024 | 0.033 | 0.018 | 0.020 | 0.019 | 0.014 | 0.021 |
| Tibooburra Al-rich chon. | 0.020 | 0.013 | 0.053 | 0.068 | 0.017 | 0.004 | 0.015 |
| Tibooburra chondrules | 0.483 | 0.607 | 0.218 | 0.312 | 0.354 | 0.242 | 0.388 |
| Tibooburra matrix | 0.466 | 0.343 | 0.668 | 0.561 | 0.601 | 0.736 | 0.568 |
|  |  |  |  |  |  |  |  |
| Nova 002 CAIs | 0.005 | 0.005 | 0.038 | 0.015 | 0.007 | 0.003 | 0.006 |
| Nova 002 AOAs | 0.029 | 0.044 | 0.043 | 0.043 | 0.025 | 0.014 | 0.028 |
| Nova 002 Al-rich chon. | 0.000 | 0.000 | 0.000 | 0.000 | 0.000 | 0.000 | 0.000 |
| Nova 002 chondrules | 0.728 | 0.784 | 0.703 | 0.674 | 0.678 | 0.631 | 0.694 |
| Nova 002 matrix | 0.238 | 0.168 | 0.216 | 0.268 | 0.290 | 0.352 | 0.272 |
|  |  |  |  |  |  |  |  |
| Bali CAIs | 0.080 | 0.064 | 0.337 | 0.239 | 0.092 | 0.038 | 0.080 |
| Bali AOAs | 0.000 | 0.000 | 0.000 | 0.000 | 0.000 | 0.000 | 0.000 |
| Bali Al-rich chon. | 0.000 | 0.000 | 0.000 | 0.000 | 0.000 | 0.000 | 0.000 |
| Bali chondrules | 0.511 | 0.641 | 0.393 | 0.320 | 0.439 | 0.325 | 0.476 |
| Bali matrix | 0.410 | 0.294 | 0.270 | 0.441 | 0.469 | 0.636 | 0.444 |
|  |  |  |  |  |  |  |  |
| Mokoia CAIs | 0.050 | 0.035 | 0.280 | 0.174 | 0.043 | 0.017 | 0.049 |
| Mokoia AOAs | 0.021 | 0.030 | 0.023 | 0.022 | 0.019 | 0.013 | 0.020 |
| Mokoia Al-rich chon. | 0.000 | 0.000 | 0.000 | 0.000 | 0.000 | 0.000 | 0.000 |
| Mokoia chondrules | 0.406 | 0.503 | 0.359 | 0.305 | 0.376 | 0.291 | 0.380 |
| Mokoia matrix | 0.523 | 0.432 | 0.338 | 0.499 | 0.562 | 0.679 | 0.551 |
|  |  |  |  |  |  |  |  |
| Vigarano CAIs | 0.028 | 0.014 | 0.141 | 0.148 | 0.031 | 0.010 | 0.031 |
| Vigarano AOAs | 0.015 | 0.022 | 0.019 | 0.022 | 0.011 | 0.005 | 0.015 |
| Vigarano Al-rich chon. | 0.000 | 0.000 | 0.000 | 0.000 | 0.000 | 0.000 | 0.000 |
| Vigarano chondrules | 0.615 | 0.715 | 0.474 | 0.466 | 0.559 | 0.422 | 0.566 |
| Vigarano matrix | 0.342 | 0.248 | 0.365 | 0.363 | 0.398 | 0.563 | 0.388 |
|  |  |  |  |  |  |  |  |
| Leoville CAIs | 0.013 | 0.008 | 0.148 | 0.104 | 0.062 | 0.007 | 0.016 |
| Leoville AOAs | 0.042 | 0.050 | 0.047 | 0.050 | 0.044 | 0.031 | 0.040 |
| Leoville Al-rich chon. | 0.007 | 0.004 | 0.017 | 0.019 | 0.012 | 0.004 | 0.006 |
| Leoville chondrules | 0.690 | 0.762 | 0.603 | 0.619 | 0.625 | 0.535 | 0.642 |
| Leoville matrix | 0.248 | 0.175 | 0.185 | 0.208 | 0.257 | 0.423 | 0.296 |

|  | Si | Mg | Al | Ca | Ti | Fe | area |
|---|---|---|---|---|---|---|---|
| Colony CAIs | 0.019 | 0.011 | 0.082 | 0.084 | 0.059 | 0.016 | 0.021 |
| Colony AOAs | 0.043 | 0.064 | 0.052 | 0.067 | 0.051 | 0.028 | 0.044 |
| Colony Al-rich chon. | 0.006 | 0.003 | 0.015 | 0.012 | 0.010 | 0.006 | 0.007 |
| Colony chondrules | 0.604 | 0.754 | 0.494 | 0.515 | 0.525 | 0.367 | 0.541 |
| Colony matrix | 0.328 | 0.168 | 0.356 | 0.322 | 0.355 | 0.583 | 0.386 |
| Kainsaz CAIs | 0.022 | 0.011 | 0.119 | 0.092 | 0.076 | 0.013 | 0.025 |
| KainsazAOAs | 0.043 | 0.059 | 0.068 | 0.058 | 0.050 | 0.024 | 0.044 |
| Kainsaz Al-rich chon. | 0.006 | 0.003 | 0.011 | 0.015 | 0.008 | 0.002 | 0.004 |
| Kainsaz chondrules | 0.611 | 0.684 | 0.428 | 0.533 | 0.501 | 0.387 | 0.520 |
| Kainsaz matrix | 0.319 | 0.244 | 0.375 | 0.302 | 0.365 | 0.575 | 0.407 |
| Ornans CAIs | 0.030 | 0.017 | 0.235 | 0.155 | 0.098 | 0.014 | 0.026 |
| Ornans AOAs | 0.033 | 0.038 | 0.043 | 0.036 | 0.037 | 0.028 | 0.023 |
| Ornans Al-rich chon. | 0.023 | 0.017 | 0.025 | 0.035 | 0.023 | 0.013 | 0.014 |
| Ornans chondrules | 0.643 | 0.693 | 0.503 | 0.508 | 0.554 | 0.469 | 0.418 |
| Ornans matrix | 0.271 | 0.235 | 0.193 | 0.266 | 0.288 | 0.477 | 0.517 |
| Lance CAIs | 0.023 | 0.012 | 0.142 | 0.113 | 0.082 | 0.013 | 0.025 |
| Lance AOAs | 0.021 | 0.028 | 0.032 | 0.028 | 0.022 | 0.014 | 0.022 |
| Lance Al-rich chon. | 0.002 | 0.001 | 0.003 | 0.003 | 0.002 | 0.001 | 0.001 |
| Lance chondrules | 0.561 | 0.609 | 0.465 | 0.546 | 0.471 | 0.364 | 0.468 |
| Lance matrix | 0.393 | 0.350 | 0.358 | 0.310 | 0.422 | 0.609 | 0.484 |
| Warrenton CAIs | 0.020 | 0.011 | 0.145 | 0.087 | 0.082 | 0.010 | 0.021 |
| Warrenton AOAs | 0.046 | 0.046 | 0.064 | 0.050 | 0.050 | 0.046 | 0.044 |
| Warrenton Al-rich chon. | 0.002 | 0.001 | 0.005 | 0.004 | 0.003 | 0.001 | 0.001 |
| Warrenton chondrules | 0.664 | 0.668 | 0.563 | 0.674 | 0.586 | 0.584 | 0.580 |
| Warrenton matrix | 0.269 | 0.274 | 0.223 | 0.185 | 0.279 | 0.359 | 0.354 |
| Acfer 094 CAIs | 0.024 | 0.015 | 0.242 | 0.177 | 0.111 | 0.009 | 0.025 |
| Acfer 094 AOAs | 0.040 | 0.069 | 0.055 | 0.048 | 0.041 | 0.021 | 0.038 |
| Acfer 094 Al-rich chon. | 0.000 | 0.000 | 0.000 | 0.000 | 0.000 | 0.000 | 0.000 |
| Acfer 094 chondrules | 0.453 | 0.554 | 0.357 | 0.297 | 0.349 | 0.280 | 0.372 |
| Acfer 094 matrix | 0.483 | 0.363 | 0.346 | 0.478 | 0.499 | 0.690 | 0.565 |

A different way to consider these data is to calculate the relative enrichment/depletion of elements among components. This is done by dividing the fraction of total (bulk) element counts accounted for by each clast type by the total number of pixels in each clast type (Table 8). These results are plotted in Figure 4 for CO chondrites Colony and Kainsaz combined, and for Acfer 094, and in Figure 5 for CV chondrites, combining Vigarano and Leoville (reduced), and Mokoia and Allende (oxidized). Acfer 094 differs dramatically from the CO chondrites in the Al- and Ca-enrichment of refractory inclusions. Results are the same within uncertainty for the oxidized and reduced CV chondrites.

## 4.6. Size Distribution of Components

Apparent size distributions of chondrules, AOAs, and CAIs (combined as 'inclusions', or 'clasts') were determined from the outlined element maps of each chondrite. Diameters are calculated for circles with area equivalent to measured inclusion pixel area. Figure 6 presents apparent diameter histograms for all of the Allende slabs (combined) and all the Allende polished sections (combined). The number of AOAs (n=64) in the sections is too small for those data to be useful. Mean radii across all inclusions are 0.605 ± 0.424 µm for slabs and 0.313 ± 0.322 µm for sections. This difference for Allende must be due to the difference in map resolution between slabs and sections (Table 3). Smaller inclusions are under-represented in large slabs mapped at low resolution, and were not corrected for using data collected at higher resolution on small subregions. Mean (standard deviation) diameters for inclusions are 752(507), 484(476) and 613(404) µm for AOA, CAI and chondrules in slabs, and 376(352), 160(226), and 385(333) µm for AOA, CAI and chondrules in polished sections.

Figure 7 shows for Allende the combined histogram for all slabs, and the histogram for all polished sections, plotted using the phi scale of sedimentary petrology (Folk, 1980). The same data is corrected according to the algorithm of Eisenhour (1996), at twice his bin resolution. In this case, only two of the correction algorithms are applied. The correction for transmitted light passing through standard 30 µm polished thin sections is omitted. The correction for the slab data does not shift the slab curve very much. The correction for the PS data shifts the mean size to a much smaller size, because smaller objects were sampled in the PS data. But the PS data is probably not representative of Allende due to its smaller area (Table 3; Hezel et al., 2008). The slab data is likely to be the best size distribution available for CV chondrite inclusions larger than about 4000 µm$^3$.

**Table 9:** Size data (microns, uncorrected) for *n* inclusions (CAIs, AOAs, and chondrules) in each section: mean, standard deviation, and median of diameter (*d*, µm), and skewness and kurtosis. "All CO" are values across all the inclusions in CO sections.

|  | $n$ | mean $d$ | 1σ | median $d$ | skewness | kurtosis |
|---:|---|---|---|---|---|---|
| Colony | 2502 | 73.93 | 68.40 | 55.49 | 3.4861 | 23.68 |
| Kainsaz | 3849 | 71.80 | 62.28 | 53.07 | 2.7965 | 14.44 |
| Ornans | 610 | 94.20 | 72.42 | 74.00 | 2.3218 | 7.81 |
| Lancé | 1569 | 111.42 | 82.33 | 90.33 | 2.4802 | 9.74 |
| Warrenton | 513 | 130.34 | 78.87 | 108.54 | 2.2376 | 8.67 |
| All CO | 9043 | 84.10 | 71.95 | 64.51 | 2.7401 | 13.62 |
| Acfer 094 | 825 | 60.94 | 56.83 | 42.73 | 2.8412 | 12.08 |

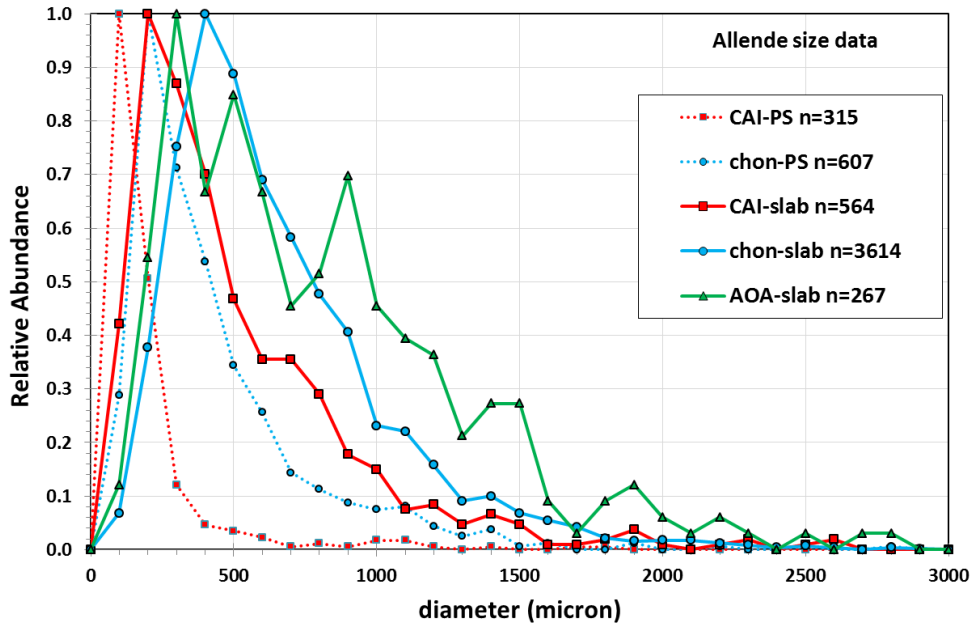

**Fig. 6:** Apparent size histogram for inclusions in slabs and polished sections of Allende (CV). The circle-equivalent diameters are plotted (*D* of circle with same area as the clast). Slabs have far more inclusions, but are sampled at coarser resolution. The mean inclusion sizes are significantly different between the slab and section populations.

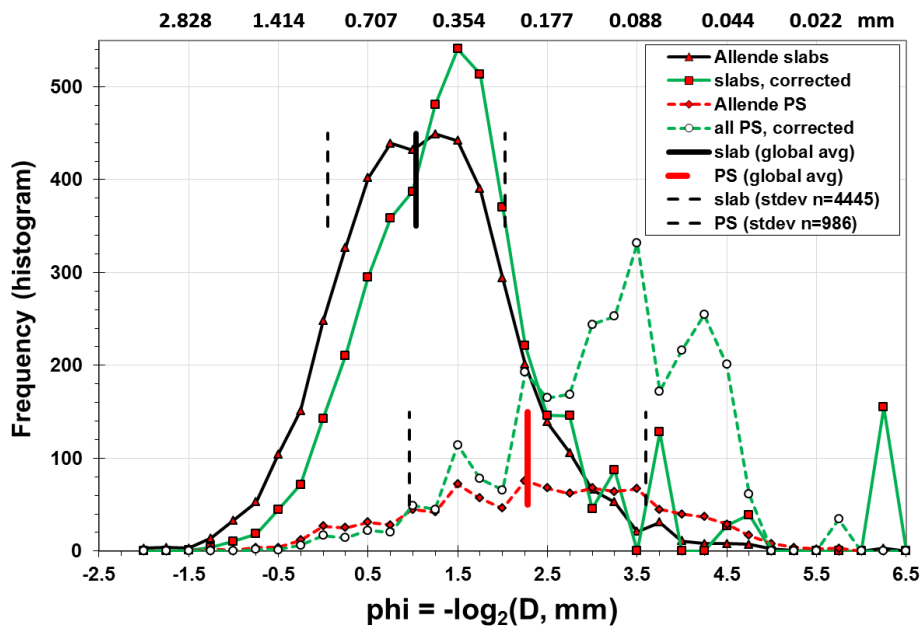

**Fig. 7:** Apparent size (phi) histogram for inclusions in all Allende slabs (n=4445) and polished sections (n=986), and the same data corrected according to the algorithm of Eisenhour (1996).

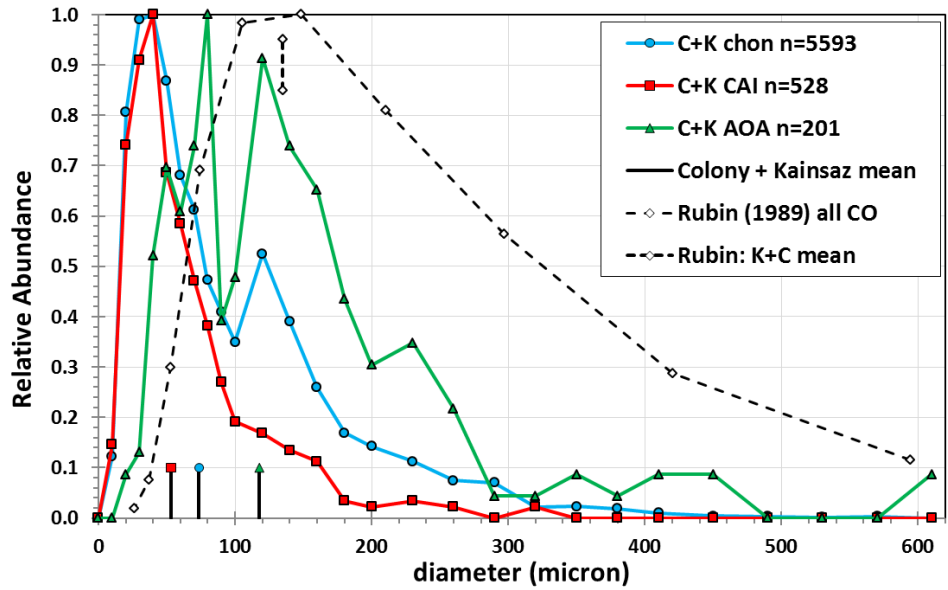

**Fig. 8:** Apparent size histogram for inclusions in Colony (CO 3.0) and Kainsaz (CO 3.2) (chondrules, CAI, AOA), uncorrected for 2D-3D. New results are binned at 0.25 phi units. Data from Rubin (1989, his Table 2) is presented for comparison.

Figure 8 shows combined size distribution results for the CO chondrites Colony (3.0) and Kainsaz (3.2), and comparison with the data reported by Rubin (1989, his Table 2). Discussion of these data is in section 5.2, below. Most inclusions fall in a narrow size distribution in the uncorrected histogram. Mean (standard deviation) diameters of circles with equivalent inclusion area are 112(89), 53(41), and 74(83) µm for AOA, CAI, and chondrules, respectively. The mean (standard deviation) diameter of inclusions across all the CO sections ($n = 9012$) is 84.16(72.02) µm. Mean diameters increase with petrologic grade from Colony (3.0) to Warrenton (3.7) (Table 9). We also present (Fig. 9) the apparent size distribution of inclusions in ungrouped Acfer 094 (C2 ungr). These data differ slightly from the CO chondrites, but display a similar hint of bimodality, particularly in the AOA size distribution.

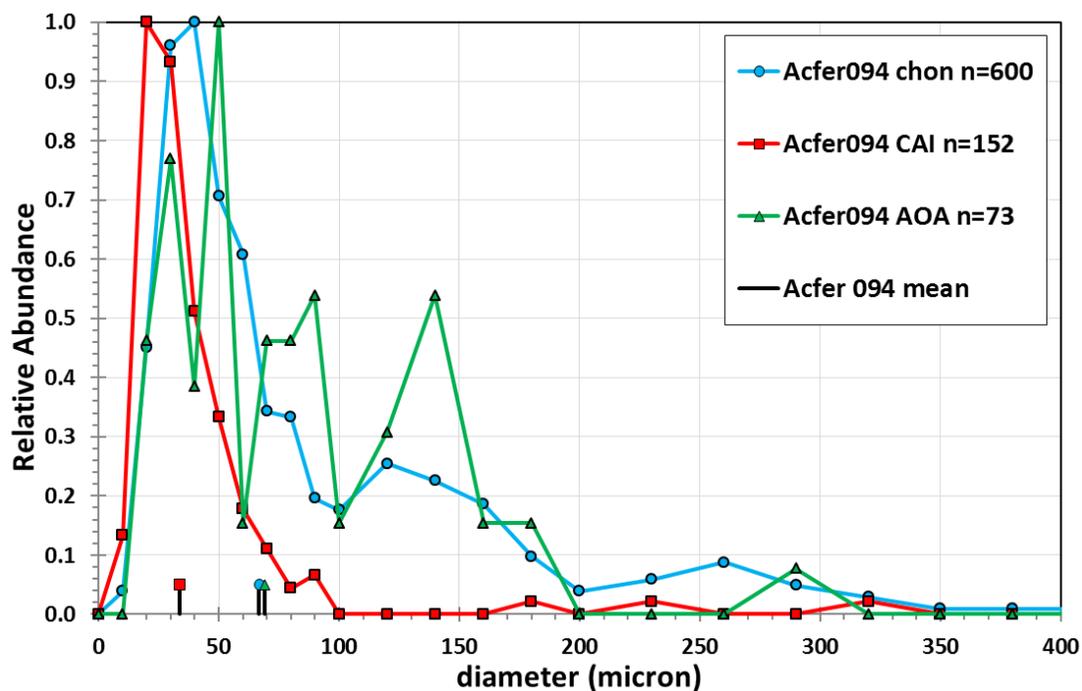

**Fig 9:** Apparent size histogram for inclusions in Acfer 094 (chondrules, CAI, AOA), uncorrected for 2D-3D.

## 5. DISCUSSION

Modern techniques have allowed us to revisit McSween's (1977a, b) pioneering measurements of relative abundances of components in CO and CV chondrites. We have covered many cm$^2$ in slabs and in sections using x-ray element mapping and manual segmentation followed by image analysis. A data set this large has not been reported previously, and it presents powerful opportunities for data mining. We have generated a selected set of statistics not only on modal abundances of objects, but also on relative element abundances in those inclusions and matrix, and size distributions of inclusions. We possess but have not yet exploited the ability to determine modal mineral abundances in each of the inclusions examined.

### 5.1. Inclusion and matrix abundance

Differences between the present results and earlier work may be assumed to be due to methodological differences. For example, most workers have grouped the accretionary rims of chondrules with matrix. However, even presumably similar methods yield variant results. For example, Rubin et al. (1985) reported their own and McSween's (1977a; personal communication) results for Colony (Table 6). Grossman and Brearley (2005) used x-ray mapping to segment (distinguish boundaries of) matrix (matrix/inclusion ratios are recalculated from their vol% matrix in Table 6). In transmitted light, the opacity of matrix is its primary distinguishing characteristic. One might expect optical point counting to over-count matrix, relative to x-ray mapping in which some matrix may be included as part of clasts. This is apparent in comparing matrix/inclusion ratios from Grossman and Brearley's (2005) automated matrix identification to McSween's (1977a and cited in Rubin et al. 1985) point counts, but is not the case in the present work that is based on manual recognition of clast boundaries in stacked elemental x-ray maps (Sec. 3.3.1).

**5.2. Inclusion size**

It is apparent from Figure 7 that the Eisenhour (1996) correction transforms log-normal distributions to differently shaped probability functions. Eisenhour (1996) noted this, and concluded that disaggregated chondrules appear to fit a Weibull function (Weibull, 1951). He suggested that a non-zero minimum clast size is possible with such a distribution (but cf., Bigolski et al. 2014). Teitler et al. (2010) demonstrated that chondrule size distributions are *not* described by a Weibull distribution. It is not clear whether application of the Eisenhour (1996) correction to our data is appropriate, because it has not been shown that the Eisenhour (1996) correction actually reproduces true 3D inclusion size distributions in chondrites (Friedrich et al. 2014, 2015). A key observation is the different size distribution measured for CV and CO chondrites (cf., Weisberg et al., 2006). The CV and CO chondrites sample different populations of chondrules and CAIs (e.g., Jones, 2012). For example, almost no igneous (Type B) CAIs are observed in CO, but Type B CAIs are much more abundant in CV chondrites (Table 7).

Size and density of inclusions are the vital parameters for dynamical model calculations (e.g., Cuzzi and Weidenschilling, 2005; Weidenschilling and Cuzzi, 2005; Brauer et al., 2007; Johansen et al., 2015). Earlier workers have quantified the radii and densities of small numbers of chondrules separated from chondrites (Paque and Cuzzi, 1997; Teitler et al., 2010), or quantified clast size and density by tomography (Kuebler et al., 1999, for OC inclusions). Previous work on sizes of chondrules in all chondrite groups is summarized in Friedrich et al. (2014). Our mean size (corrected) for CO chondrite clasts is about half that determined by Rubin using a calibrated reticule in a petrographic microscope objective (1989, his Table 2). Rubin (1989) reported a mean diameter of 148 micrometers (+132/-70) for CO chondrite chondrules only, uncorrected. In Figure 8, his data are plotted, with his mean size. Comparisons of the data suggest that the objects that Rubin (1989) determined to be chondrules is a subset of what we recognize here as chondrules. There is a second peak in our curves for both AOAs and chondrules, corresponding nearly to Rubin's mean diameter. Rubin (1989) explicitly did not measure "fragments", arguing that chondrule fragmentation is a parent body impact gardening process. However, all of the CO chondrites measured in this work are shock stage S1 (Scott et al. 1992). The CO chondrites (e.g., Fig. 1a), and indeed, all the carbonaceous chondrites considered here (Fig. 1), display a continuum of shapes of which nicely rounded chondrules are one extreme. Fragmentation might also be expected to produce a continuous size distribution, rather than the bimodal distribution suggested by our results. The reason for this bimodality (Fig. 8) is unknown.

**5.3. CAI contribution to refractory element enrichment**

The refractory inclusions (CAI + AOA) contribute 20% (1σ=5%) of the Al, 15% (1σ=2%) of the Ca, and 12% (1σ=1%) of the Ti to the average CO chondrite bulk compositions (Fig. 4, Table 8). Depending on how one normalizes the CO chondrite bulk compositions for comparison to CI, it might be possible to use this data to answer the question: Is some fraction of the CAI abundance in CO (and CV) chondrites responsible for the excess (relative to CI) refractory elements in CO and particularly in CV chondrites (Table 1)? That is, is there a non-complementary (i.e., super-chondritic) CAI addition to these accretionary bodies that is not sampled by CI, CM, CR and OC chondrites, and, if so, how much? The present data hint at a solution to this problem, but the data are not of sufficient precision to justify a quantitative answer to the question. Preliminary results for REE in CO chondrite inclusions suggest that no such non-complementary CAI component exists (Crapster-Pregont and Ebel, 2014b). The bulk refractory element abundances in CO and CV chondrites differ strongly, despite their nearly similar CAI abundances (Tables 5, 6), suggesting that CAI enrichment may not be the direct cause of refractory element enrichment of CVs.

Hezel et al. (2008) concluded that CAIs "contributed little to the bulk chondrite refractory element abundances", and also that they "formed in a separate nebular region from chondrules and matrix". May et al. (1999) reach a similar conclusion, and make the interesting point that the aerodynamic properties of mostly 'fluffy' CAIs must be significantly different than the aerodynamics of compact, melted, spherical (Type B) CAIs. The rarity of Type B CAIs (Table 7) makes such a conclusion difficult to test. The present results do not support the aerodynamic sorting of chondrite components, which would be likely to also sort bulk compositions, resulting in non-chondritic abundances of major elements in chondrites. The refractory inclusions (CAI + AOA) contribute 18% (1σ=11%) of the Al, 14% (1σ=7%) of the Ca, and 6% (1σ=3%) of the Ti to the average CV chondrite bulk compositions (Fig. 5, Table 8). This is within range of the calculation by Rubin (2011) that refractory inclusions contain 21% of the Ca in Allende. Our result for CAIs in Allende (12% of Ca) is very close to the 14% calculated by Rubin (2011).

**5.4. Ice in reduced vs. oxidized CV chondrites?**

An unexpected result, not anticipated by McSween (1977b, c), is the large difference in clast/matrix ratio between the oxidized and reduced subgroups of CV chondrites (Fig. 2). Bali is the exception, however Bali exhibits "extreme heterogeneity of alteration" (Scott et al., 1997). Earlier, we reported the high abundance of matrix in Allende (Ebel et al., 2008a). Here we refine those results and confirm similar abundances in Mokoia (OxA) and Tibooburra (presumably also OxA subtype). Dyl et al. (2014) demonstrate that matrix in Vigarano (CV-R), with much lower porosity (Table 2) is less homogenized than matrix in Allende (CV-OxA) at cm-mm scales. We momentarily set aside data for heavily weathered Leoville, and for Tibooburra and Bali for which porosity has not been reported. We rely on our analysis to establish that Nova 002 is a reduced CV chondrite.

Matrix is, on a fine scale (nm-μm), the most heterogeneous component of chondritic meteorites. However, on a coarse scale (cm), matrix is homogeneous.

Evidence for five different primary components of chondrite matrix was described by Huss et al. (2005): 1) average dust from the Sun's parent molecular cloud, 2) material that evaporated from inclusions and recondensed onto matrix, 3) inclusion fragments, and locally evaporated molecular cloud material, either 4) fully or 5) partially recondensed as or on fine grained dust. Here, we provide circumstantial evidence that the excess pore space in oxidized CV contained ice grains as an additional primary accreted component.

Macke et al. (2011) measured chondrite porosity by He pycnometry. The mean porosity of Allende and Mokoia (CV-OxA) is 24.8 vol% and for Nova 002 and Vigarano (CV-R) it is 7.2% (Table 2). The average matrix vol% for these CV-OxA is 55.9% and for the two CV-R it is 37.6%. The 17.7% difference in porosity closely matches the 18.3% difference in matrix vol%. We can, therefore, consider the 'grain volume' in matrix to be nearly the same in both the oxidized and reduced CV, under the assumption that the porosity in all CV is primarily in the matrix, or that the porosity in inclusions is the same among CV chondrites, and the *difference* resides in the matrix. The presence of an ice grain component volumetrically equivalent to the presently observed porosity difference is consistent with our findings on the relative abundances of matrix in oxidized (OxA) and reduced subgroups. We have not corrected for the fact that matrix grains are coarser in CV-Ox than in CV-R (Krot et al., 1995), and that the molar volume of fayalite is 4% larger than that of forsterite (Yoder and Sahama, 1957; Fujii, 1960). Because the primary mineralogy of CV chondrite matrix remains uncertain (Brearley and Jones, 1998, p.220), we cannot go much further with this analysis.

Our results lend support to the idea that ices are transported into chondrite parent bodies with matrix (Bunch and Chang, 1980) and/or with fine-grained rims on chondrules (Bigolski et al., 2014). Grimm and McSween (1989) considered 0.4 an upper limit on the ice volume fraction in CM and CI carbonaceous chondrite parent bodies. Thus the enhanced aqueous alteration in the oxidized CVs may be attributed to the accretion of more ice-bearing matrix into the oxidized CV than into the reduced CV. This remains circumstantial evidence, as small ice grains present in the parent rock are long since gone (Ebel et. al., 2014). Our results are consistent with Palme and Pack's (2008) suggestion that ice grains are associated with unaltered presolar grains. If ice grains were major original components of matrix, along with presolar grains and organics, then the chondrites with higher matrix abundances would have had more original water to promote oxidation (hence CV-Ox; MacPherson and Krot, 2014). Allende (CV-OxA) does not exhibit cm-scale effects of aqueous alteration (Stracke et al., 2012; Dyl et al., 2014), so all of these effects must be represented by highly local mineralogy.

It may seem counterintuitive that the porosity of Allende is high, even though Allende is thought to have been heated to the highest temperature of the CV chondrites studied here. MacPherson and Krot (2014) argue that $^{26}$Al was the major heating source during aqueous alteration of asteroids. Many researchers have documented the difficulty in disentangling aqueous and thermal processes in CV chondrites (e.g., Krot et al., 1995). The apparently undisturbed U-Pb systematics of Allende Type B CAIs (Connelly et al., 2012), and extreme alteration of opaque minerals in chondrules (Fu et al., 2014), argue for heterogeneous effects of alteration on different types of Allende inclusions. Whether heating and aqueous alteration would also require compaction, reducing porosity, requires knowledge of the original properties of the CV parent body or bodies. MacPherson and Krot (2014) invoke differential compaction and expulsion of water, rather than heterogeneous original abundances of ices, to explain the differences in CV alteration styles. We note, however, that the CV chondrites studied here are shock stage S1 to S1-S2 (Table 2), so compaction by impact gardening of these rocks was not extensive (cf., Rubin, 2012).

Weisberg et al. (1997) identified a suite of characteristics that distinguish the OxA and OxB subgroups of CV3 chondrites. Our results (Table 4; Fig. 2) differ from the matrix/chondrule ratios estimated in their preliminary study. They reported low matrix/chondrule ratios for the reduced subgroup, slightly higher, 0.5-0.6 for the OxA subgroup, and 0.6-0.7 for the OxB subgroup.

The CO chondrites do not exhibit any correlation between petrologic grade and matrix/inclusion ratio (Fig. 2). Nor is there any trend in measured porosity with petrologic grade in CO chondrites (Macke et al., 2011). This result is consistent with the thermal, rather than aqueous alteration experienced by the CO chondrites (Keller and Buseck, 1990). However, the CO with the highest observed porosity (Table 2), Ornans, also has the highest matrix/inclusion ratio of all the CO chondrites studied here. An alternative interpretation of higher matrix abundance with petrologic grade is that of McSween (1979). He proposed that apparent matrix contents observed in CM chondrites may not be primary, but instead result from optical effects due to aqueous alteration. X-ray maps show AOAs and "fluffy" CAIs being "blending into" or being absorbed into matrix in more altered CV chondrites (e.g., Allende, Grosnaja). In Allende we observe that at coarse resolution (slabs) some finer-grained non-igneous inclusions (fluffy CAIs, AOAs) appear to be hidden, obscured by aqueous and thermal alteration.

**5.5. Element Distributions: Mg - Si - Fe**

*5.5.1. Metamorphic effects in CO and CV chondrites*

The Mg-Si relations between inclusions and matrix among CO chondrites from 3.0 to 3.7 are shown in Figure 10, with data normalized for the assumption of uniform bulk CO compositions across the population. With increasing petrologic grade the Mg/Si ratios of clasts and matrix approach the bulk (solar) values, although Ornans 3.4 and Lancé 3.5 deviate slightly from this pattern. No such correlation of Mg/Si ratio with petrologic grade is observed in the CV chondrites (Fig. 11).

33333333

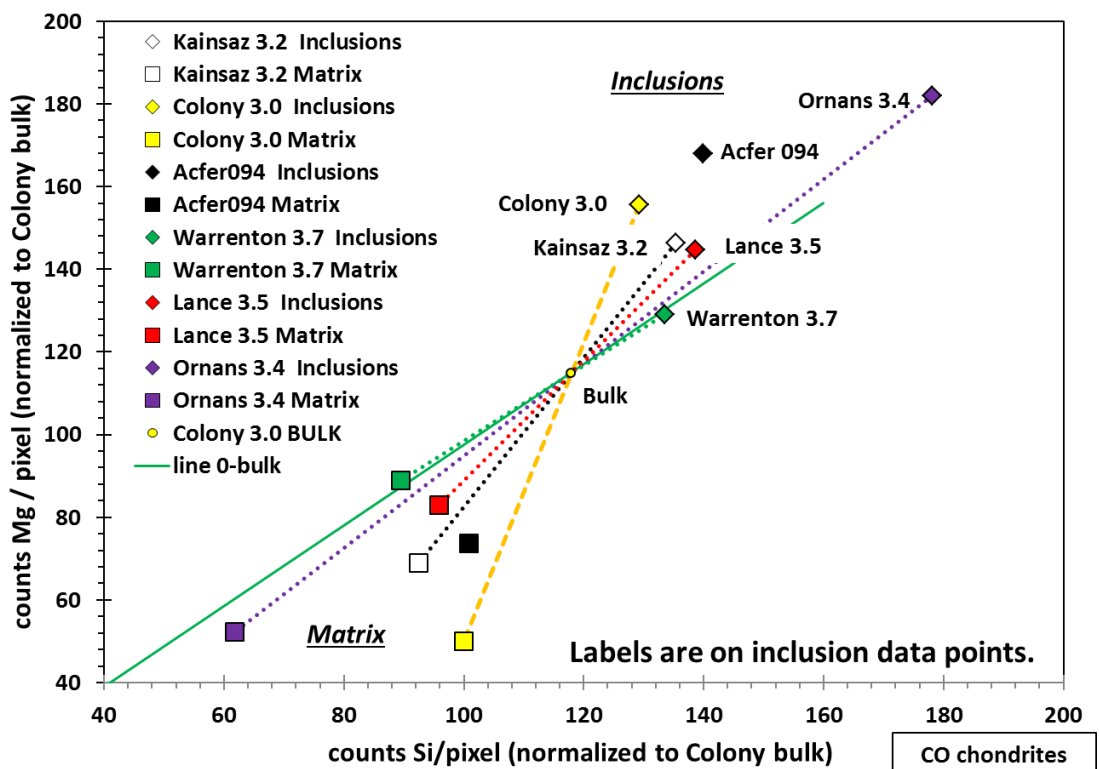

**Fig. 10:** Mg-Si relations in CO chondrites, normalized to Colony bulk composition. Solid green line joins bulk ratio to origin. Mg/Si ratios approach solar with increasing petrologic grade, but Si remains enriched in inclusions.

In all chondrites, it is well known that with increasing petrologic grade, Fe and Mg are increasingly redistributed among inclusions and matrix (Krot et al., 1995; Brearley and Jones, 1998; Grossman and Brearley, 2005). Figure 10 illustrates that by CO grade 3.7 (Warrenton) the Fe-Mg exchange of inclusions and matrix is essentially complete, such that Mg/Si of inclusions and matrix match the solar (bulk CO) Mg/Si ratio. No such wholesale exchange is observed in the CV chondrites (Fig. 11). In the CO chondrites the mobile element is Mg, rather than Si (Fig 10).

Figure 12 illustrates the changes in Fe and Mg with petrologic grade in CO chondrites. In the highest grade CO Warrenton (3.7), and in CO 3.4 Ornans (but not in CO 3.5 Lancé), Fe is nearly equilibrated between inclusions and matrix. A reasonable interpretation of the data in Figures 10 and 12 is that Fe-Mg exchange between inclusions and matrix becomes more complete with increasing petrologic grade. Indeed, this is qualitatively understood through petrological examination of CO chondrites (Brearley and Jones, 1998). Here, we are able to quantitatively demonstrate these changes in the chemical balance among components. Note also that for Warrenton (CO 3.7) the differences in Si (Fig. 10) and Mg (Fig. 12) between matrix and inclusions are smaller than for the lower petrologic grade chondrites. This suggests that once solar Mg/Si is reached and Fe equilibrates between inclusions and matrix, the next stage of alteration involves equilibration of Si and equilibration of Mg such that both converge on the bulk value. The CV chondrites show no such systematic variation with petrologic grade.

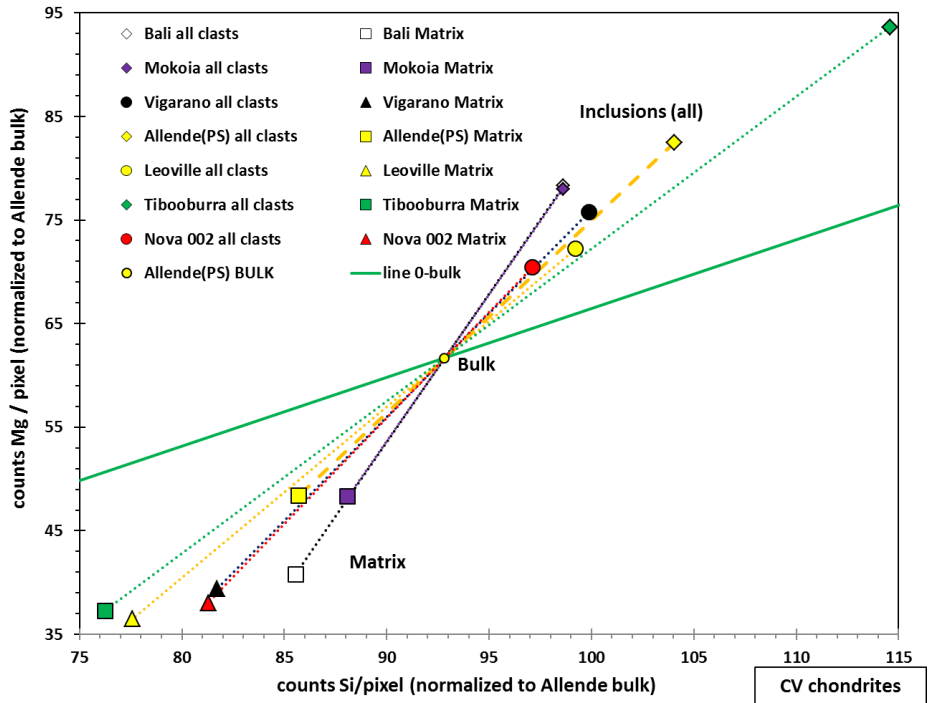

**Fig. 11:** Mg-Si relations in CV chondrites, normalized to Allende bulk composition. Solid green line joins bulk ratio to origin.

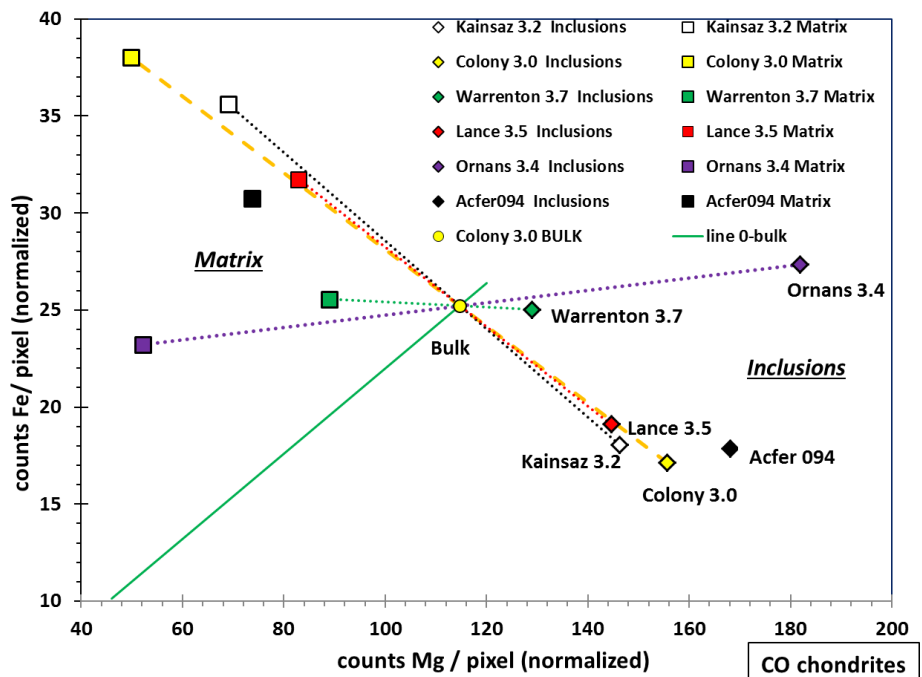

**Fig. 12:** Mg-Fe relations in CO chondrites, normalized to Colony bulk composition. Solid green line joins bulk ratio to origin.

Colony (CO 3.0) may have reached 200°C, with minimal aqueous alteration. Kainsaz (3.2) lacks phyllosilicates but may have reached ~300°C, and CO 3.7 reached ~600°C (Huss et al., 2006). Warrenton (3.7) has no matrix phyllosilicates, but Lancé (3.5) does (Keller and Buseck, 1990). The fine-grained olivine in CO chondrite matrix becomes completely equilibrated from type 3.0 to 3.7 (Brearley and Jones, 1998, their Fig. 144). Our results (Fig. 12) show that the Fe/Mg ratios of inclusions and matrix generally approach the bulk Fe/Mg ratio with increasing petrologic grade.

*5.5.2. Complementarity of Fe-Mg-Si between inclusions and matrix*

"Some of it plus the rest of it equals all of it" is a law that obscures the fact that "all of it" is chondritic within a factor of 1.5 for CV and 1.1 for CO (Weisberg et al., 2006, their Fig. 3; Palme et al., 2015, their Figs. 2, 3). This is observed for a huge number of meteorites with wide variations in inclusion/matrix ratios and relative abundances of inclusions (e.g., CAI/chondrule, chondrule types). Here, we demonstrate that the "law" applies to CO and CV chondrites, which have very different mean sizes of inclusions. The CI composition has "special significance" in this context (Huss et al., 2005, p.705). Why should chondrites be chondritic? The answer must supply a first order constraint on the origin of chondritic meteorites.

The complementary distribution of major elements among both clasts and matrix is apparent in the Mg-Si-Fe data on CO and CV chondrites. The fact that the present data include vastly more clasts and matrix pixel analyses than previous work lends credence to the constraint of matrix-chondrule "complementarity" on chondrule formation (Palme, 1992; Hezel and Palme, 2008, 2010; Palme et al., 2015). Figures 10 and 11 illustrate that inclusions and matrix have very different Mg and Si compositions, even among the CO chondrites. Yet the bulk Mg and Si compositions of CO and CV chondrites are nearly identical (Table 1; Hezel and Palme, 2010, their Table 2). This complementary relationship between inclusions and matrix from CO 3.0 to 3.7 does not appear to be *caused* by thermal or aqueous alteration after chondrule formation (Zanda et al., 2006). It is, rather, persistent *despite* such alteration. Even as alteration moves inclusion and matrix Fe/Mg ratios toward the bulk CO ratio (Fig. 12), Si does not change in either component. Thus complementarity is a primary nebular effect, and persists despite alteration.

Figure 3 illustrates Mg-Si relationships in detail for CO chondrites Colony and Kainsaz (see Fig. S3 for Ornans and Lancé). Figure 13 shows Mg-Si in Allende (Fig. S4 for Tibooburra and Nova 002, Fig. S5 for Mokoia and Vigarano). Figure 14 shows Fe-Mg in Mokoia (see Fig. S6 for Allende and Vigarano, Fig. S7 for Tibooburra and Nova 002). Figures S8-S9 illustrate Mg-Fe relationships for CO chondrites. Figure S10 illustrates Ti -Al relationships for Ornans (CO 3.4) and Colony (CO 3.0). Figure S11 illustrates Ca-Al relationships for Allende and Colony. Results for Acfer 094 (C2-ungr) are presented in Figure S12, and show similar broad chemical variability among inclusions.

These plots demonstrate that the bulk composition of any single inclusion, of any type, yields absolutely no clues as to its host chondrite's class, as also demonstrated by earlier work (Hezel and Palme, 2010 and references therein). The composition range among all inclusions is huge. Yet the inclusions combine to produce chondrites with chondritic major element compositions indicative of membership in a single chemical group. Of course, in these graphs there must be an exact coincidence of matrix and mean chondrule compositions along mixing lines resulting in bulk composition (all pixels). It is the additional fact that in these bulk compositions the major element ratios are chondritic (Table 1) that makes the complementary nature of inclusions and matrix so striking.

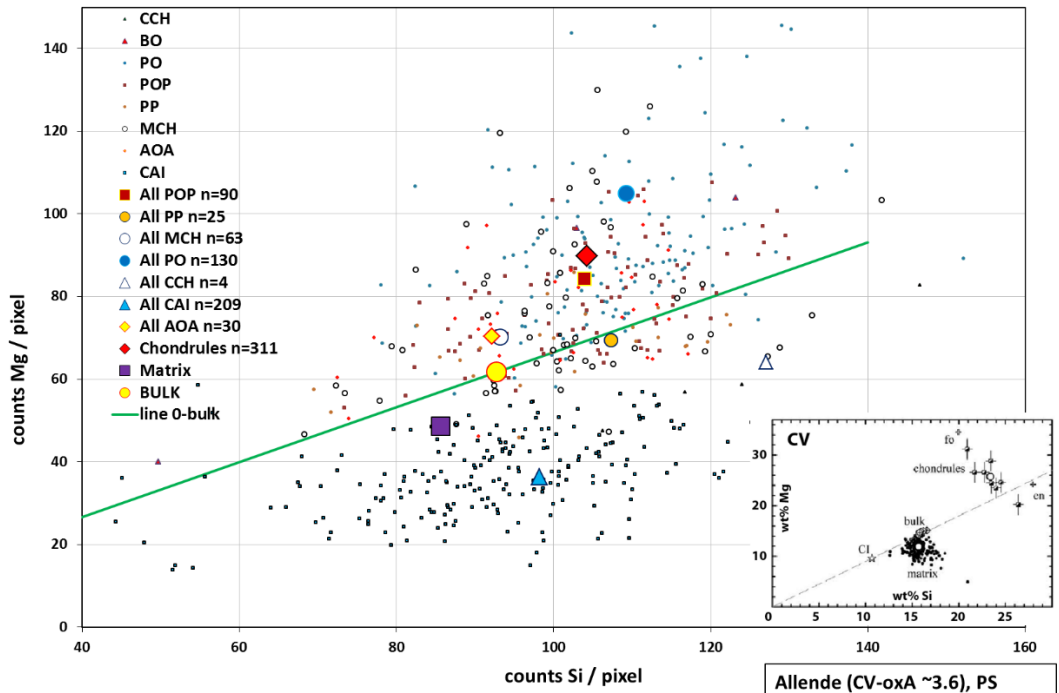

Fig. 13: Per pixel counts of Mg and Si for clasts, mean clasts, matrix, and bulk meteorite for Allende (CV~3.6) polished sections. Inset is from Hezel and Palme (2010, their Fig. 1).

The Mg-Si relations among chondrules of various types (PP, POP, etc.) are also indicated in Figures 3 and 13 and Figures S3-S5. The chondrules are, in aggregate, complementary to each other in producing the total chondrule element budget. In this work the division of chondrules into subtypes is subject to observational biases, however, the general distribution of pyroxene-rich (PP), olivine-pyroxene (POP), and olivine-rich (PO) chondrules supports the idea that chondrules themselves are complementary in yielding a bulk chondrule average that is complementary to matrix. Plots in Fe-Si space (Fig. 14; Figs. S6-S9) illustrate very similar, complementary relationships among the various clast types, bulk, and matrix.

If inclusions formed from a separate reservoir from matrix, then, given their large variation in composition, the mixing from that reservoir would have to be sufficiently stochastic that inclusions combine to make exactly the total inclusion composition that complements matrix. Among and within the different chondrite groups (here, CO and CV), that total inclusion composition must differ so as to exactly balance matrix, because inclusion/matrix ratios vary among and within the chondrite groups (Fig. 2). It is a simpler hypothesis, that material of chondritic bulk composition in one local nebular region experienced the inclusion formation process to greater or lesser extent, and that resulting inclusions and matrix combined locally to produce chondritic meteorite parent bodies.

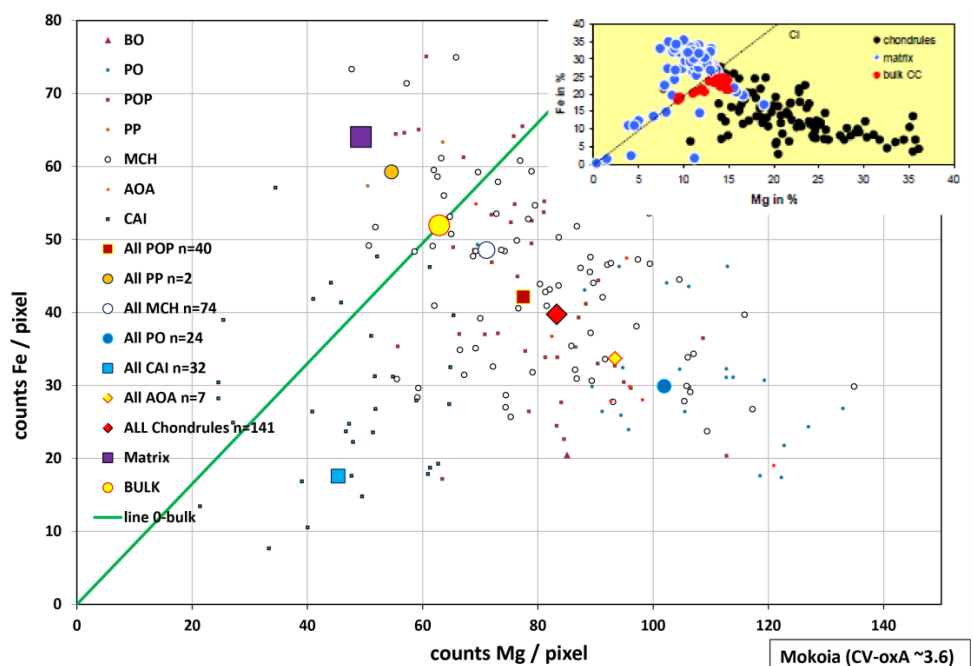

Fig. 14: Per pixel counts of Mg and Fe for clasts, mean clasts, matrix, and bulk meteorite for Mokoia (CV~3.6). Inset is from Palme and Hezel (2011, their Fig. 2).

## 5.6. Element Distributions: Al - Ti - Ca

### 5.6.1. *Complementarity of Al-Ti-Ca between inclusions and matrix*

The interpretation of Al-Ti-Ca distributions is more difficult because these elements are much less abundant than Mg, Si and Fe. Klerner and Palme (2000) demonstrated with a limited set of chondrules ($n = 72$) in Renazzo (CR) a matrix to chondrule Al ratio of 44:55, but only 12:78 for Ti, with a matrix/chondrule modal abundance ratio of 0.68. They found that mean Ca/Al is chondritic in chondrules, matrix and bulk, so their results indicate that Ti is more fractionated into chondrules than Ca or Al. Palme et al. (2011) extended this work to Mokoia, using data from Klerner (2001) and from Jones and Schilk (2009). McSween and Richardson (1977) and Zolensky et al. (1993) observed that matrix also has a subchondritic Ti/Al ratio in CV chondrites. The CR, CV and CO are all chondritic (±10%) in bulk Ti/Al and Ca/Al.

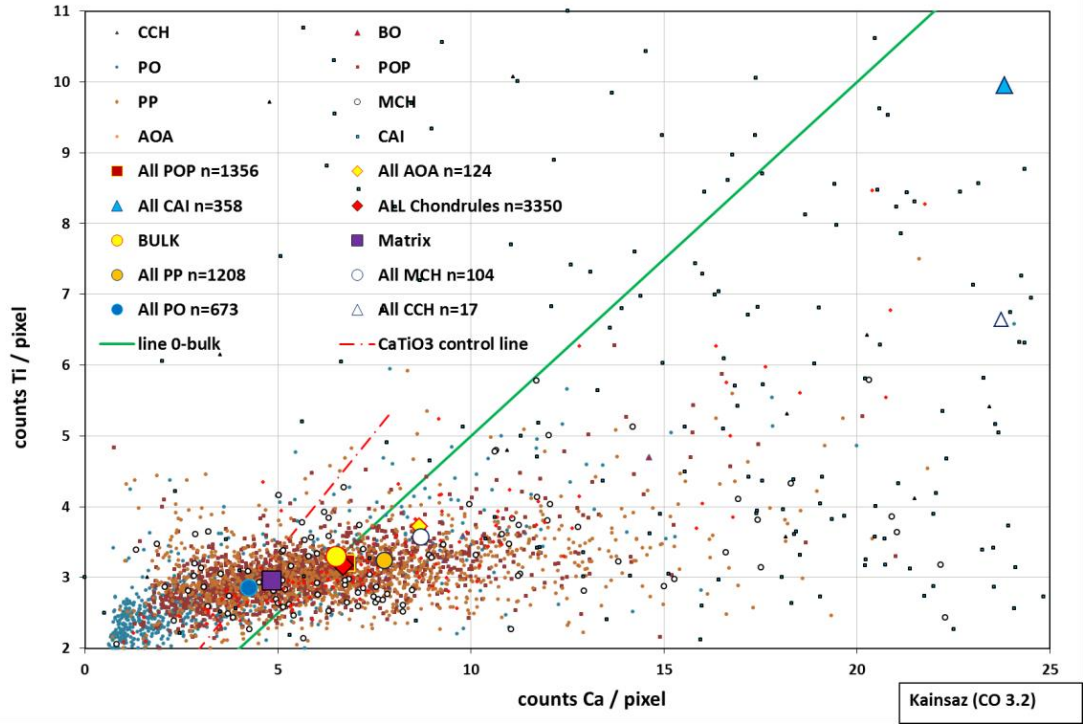

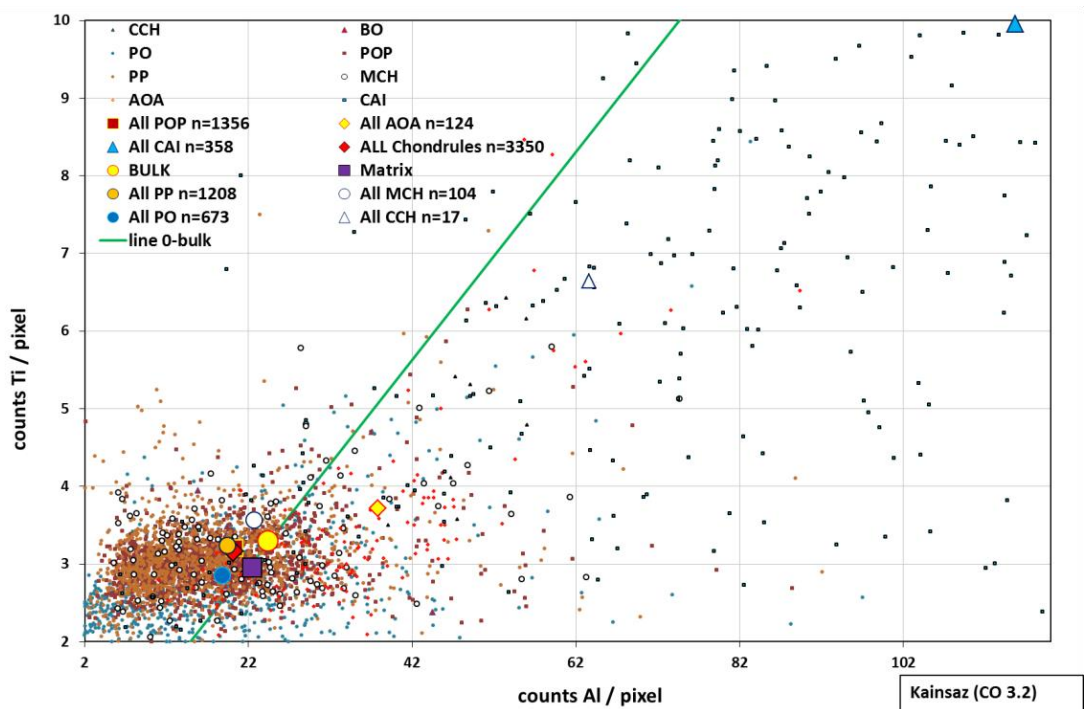

Fig. 15: Ti relative to Ca and Al compositions in Kainsaz (CO 3.2). Lines join bulk composition and origin. Upper plot shows Ti/Ca count ratios from a $128^2$ pixel map of a perovskite (CaTiO$_3$, red dash-dot) standard, normalized to conditions of Kainsaz analysis. Solid green lines join bulk compositions to origin.

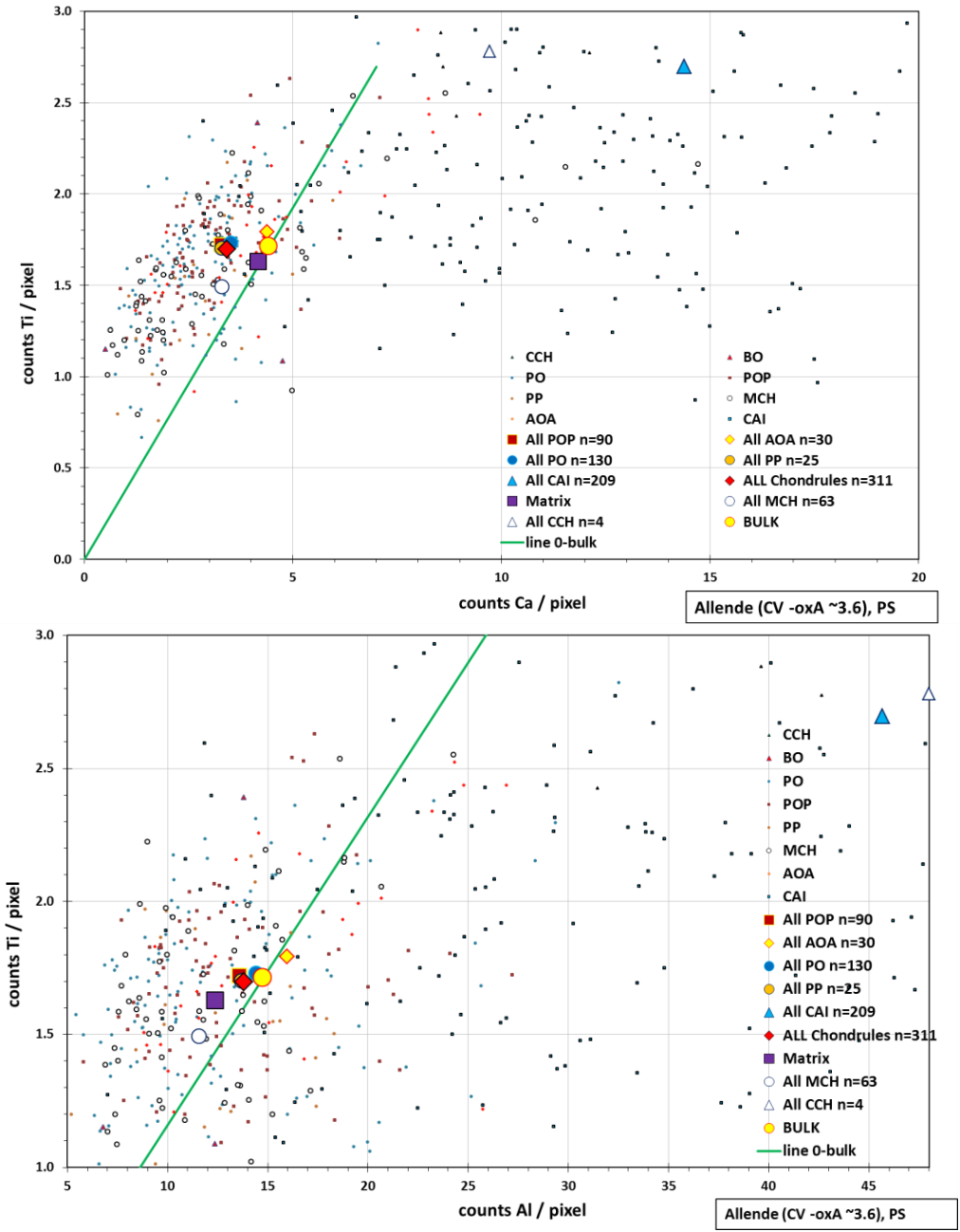

Fig. 16: Ti relative to Ca and Al compositions in Allende (CV oxA ~3.6) polished sections AL2ps5 and AL2ps9. Solid green lines join bulk compositions to origin.

We find that Ca/Al is near bulk (solar) in mean inclusions and matrix in both the CO and CV chondrites. In Figure 15, Ti (counts/pxl) in all inclusions and matrix, and mean values over inclusion types, are plotted against Ca and Al (counts/pxl) for Kainsaz (CO 3.2). The ratio of counts on a mapped perovskite ($CaTiO_3$) standard is used to establish a '$CaTiO_3$ control line' (see below). Matrix has the bulk Ti/Al ratio and exceeds bulk Ti/Ca, but is depleted in Ti, Al and Ca relative to bulk and, especially, relative to CAIs and Al-rich chondrules (CCH in Figs. 15 and 16). Chondrules in Kainsaz have higher Ti/Al than either matrix or CAIs. Combined data from Allende (CV) polished sections AL2ps5 and AL2ps9 (Fig. 16) illustrate subtly different Ca-Al-Ti relations compared to Kainsaz (CO). Again, the CAIs and Al-rich chondrules have significantly lower Ti/Ca and Ti/Al than other components. But the matrix Ti/Al is above bulk, while matrix Ti/Ca is the same as the bulk. In Allende, the chondrules differ more from bulk than the chondrules in Kainsaz. The other CV chondrites analyzed here, except Tibooburra, resemble Allende in these metrics.

Previous work (Palme and Klerner 2000; cf. Klerner and Palme, 2000) concluded that the Al-Ti distribution among CR and CV chondrite components cannot be explained by models that remelt matrix to form chondrules (e.g., Anders 1964), but instead requires perovskite in chondrule precursors. Our results for CV chondrites (Fig. 16), but not for CO (Fig. 15) appear to be consistent with this interpretation. The location of CAIs (mean) in Figures 15 and 16 indicates highly sub-chondritic Ti/Al in CO and CV CAIs. If the REE were first concentrated at high temperature in $CaTiO_3$ perovskite (Lodders, 2003), then much of that perovskite did not end up in the CAIs we have measured.

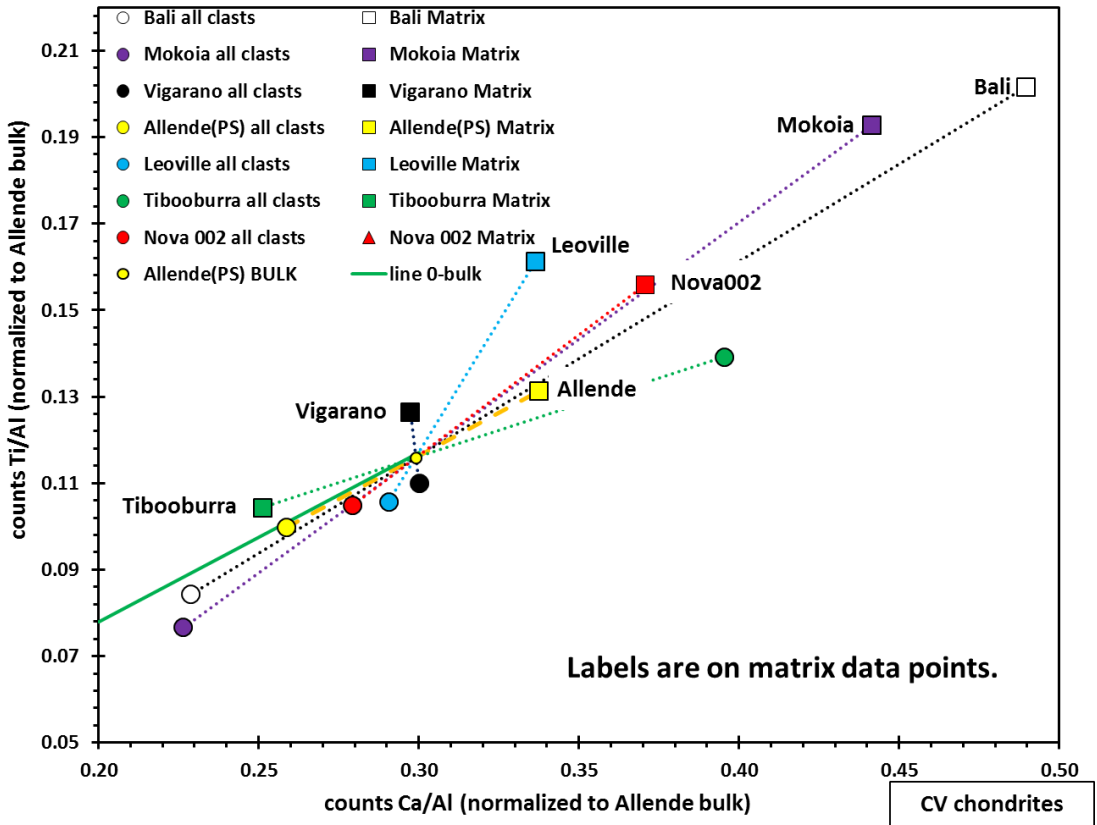

Fig. 17: Ti/Al relative to Ca/Al composition in CV chondrites, all normalized to Allende polished section bulk composition (counts/pixel).

Furthermore, without the significant Al content of CAIs and AOAs, the bulk composition of Kainsaz would be depleted in Al relative to Ti, when normalized to solar (chondritic) ratios. This strongly suggests a complementary relationship among all the inclusions (CAIs, AOAs, chondrules) and matrix, not simply between matrix and chondrules as concluded by Palme and Klerner (2000). Given that the chondrule/CAI modal ratios in Colony and Kainsaz are 25.24 and 17.92, respectively (Table 6), the mean chondrule Ti and Al contents need only move a small distance from the bulk, and above the Ti/Al line, to offset the CAI compositions (Fig. 15). The fact that the chondrules do just that indicates that the chondrules and CAIs are complementary to each other in this CO chondrite. Similar plots for the other CO chondrites show the same relationships.

These relationships are apparent in Figure 17, where Ti/Al is plotted against Ca/Al for all the CV chondrites, normalized to Allende bulk composition. Note that all inclusions are summed in Figures 17 and 18. All the CV studied here have matrix Ti/Al and Ca/Al well above bulk, and inclusions well below bulk, except Tibooburra, where the reverse is observed. Hezel and Palme (2008) observed a similar opposite relationship of Ca/Al between matrix and chondrules in Allende, a "normal CV", compared to Yamato 86751 (Murakami and Ikeda, 1994), which appears to be like Tibooburra. The "normal CV" have low Al in matrix and high Al in inclusions, whereas "abnormal CV" Tibooburra (and Y-86751) have high Al in matrix, and low Al in inclusions. Hezel and Palme (2008) pointed out the near identical bulk compositions of the CV chondrites Allende and Y-86751, and noted "common" tiny (≤ 1 μm) Al-spinel grains in the matrix of Y-86751 (Murakami and Ikeda, 1994) but not in Allende. We have not examined Tibooburra for similar spinels. However, our results reinforce the arguments presented by Hezel and Palme (2008), who summarized four possible explanations, and concluded that a "primary chemical connection" best explained Ca-Al systematics, with chondrules and matrix forming from the same reservoir.

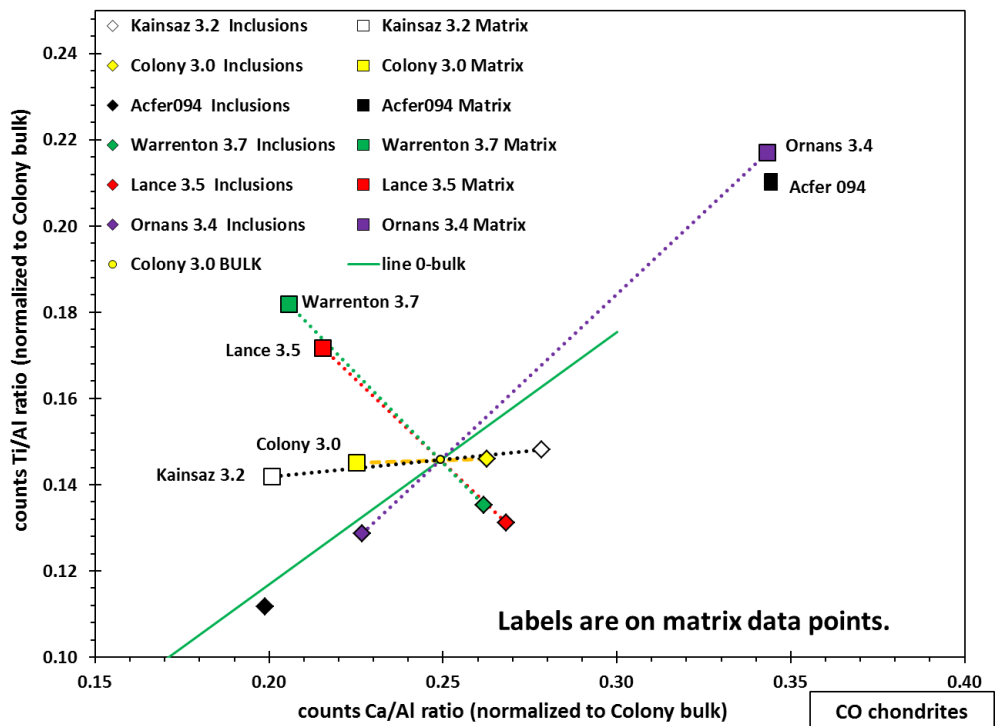

Fig. 18: Ti/Al relative to Ca/Al composition in CO chondrites, all normalized to Colony polished section bulk composition (counts/pixel).

## 5.7. Constraints on astrophysical models

### 5.7.1. *The problem*

Mechanisms for chondrule formation and meteorite parent body accretion must predict inclusion size distributions, abundances, and chemical compositions of all components and of the bulk rocks. We have provided further information on these parameters. Competing models can be eliminated if inclusions and matrix in chondrite classes are shown to originate from a single reservoir, rather than from separate reservoirs. A great deal has been written on both counts (e.g., Anders, 1964; Grossman and Wasson, 1982; Palme et al., 1992; Alexander, 2005; Zanda et al., 2006; Hezel and Palme, 2010; Palme et al., 2015).

In one class of models, differences among chondrite classes are attributed to aerodynamic sorting (e.g., due to nebula-wide turbulence, Cuzzi et al., 2001, 2003; Cuzzi and Weidenschilling, 2005; Weidenschilling and Cuzzi, 2005) or other transport phenomena in the protoplanetary disk (Shu et al., 2001; Ciesla, 2011), prior to accretion of the parent asteroids or their precursor bodies (e.g., Brauer et al., 2007). Some models of this type are intended to explain the transport of high-temperature inclusions from the innermost disk to colder regions (Brownlee, 2014).

Alternatively, inclusions may be heated further out in the disk. The most energetic process present in the early disk is the viscosity-driven accretion of the central star and concomitant outward transport of angular momentum (King and Pringle 2010; Ji and Balbus, 2013). Stellar accretion may be driven by two types of turbulent instability that dissipate energy through heating: gravitational instability and magnetorotational instability. Model heating by shocks driven by gravitational instability predicts the abundances of chondrules of various types produced for assumed initial parameters (Desch and Connolly, 2002; Chiang, 2002; Morris and Desch, 2010). Current sheet heating driven by magnetorotational instabilities (Joung et al., 2004; Hubbard et al., 2012; McNally et al., 2013) should yield similar predictions, but with more localized heating, required to preserve thermally fragile matrix. Both of these scenarios might apply well out in the disk. Shock heating (e.g., Hood, 1998; Ciesla and Hood, 2002; Desch and Connolly, 2002; Boss and Durisen, 2005) is better understood through numerical simulation than is current sheet heating (e.g., McNally et al., 2013). Both of these mechanisms are potentially consistent with the complementary nature of chondrite components.

More speculative chondrule formation models involve impacts among early planetesimals (Zook, 1981; Lugmair and Shukolyukov, 2001; Asphaug et al., 2011; Sanders and Scott, 2012), exhalations from molten planetesimal "leaky pressure cookers", or jetting of molten material in small body collisions (Keiffer, 1975; Johnson et al., 2015). All these mechanisms are expected to produce molten droplets of widely varying sizes, within an extremely narrow time window prior to planetesimal differentiation. Measurements of the actual size ranges of inclusions and size variance among chondrite types can constrain the degree of size sorting required after such proposed chondrule formation events. This class of models either requires separate reservoirs of inclusions and matrix, or a highly tuned mechanism to re-combine components to make chondritic parent bodies (i.e., preserve complementarity) that also contain presolar grains and organics in matrix-normalized abundances uniformly across chondrite classes (Huss et al., 2003).

*5.7.2. New constraints*

The strong chemical composition and abundance differences between matrix and chondrules, across chondrite groups, have led to the hypothesis of compositional "complementarity" of these components (Palme et al., 1992; Palme and Hezel, 2011; Klerner and Palme, 1999; Bland et al., 2005; Palme and Pack, 2008; Hezel and Palme 2008, 2010; Scott and Krot, 2005; Palme et al., 2015). This powerful constraint rules out separate reservoirs of chondrules and matrix in chondrite formation. Here, we have provided new data that strongly supports the hypothesis of chondrule - matrix complementarity. We have also presented evidence that supports complementarity between the various chondrule types, and between the CAIs and the rest of their host chondrites. This data is collected using techniques different from that of, e.g., Hezel and Palme (2008, 2010). Combined with previous work (e.g., Ebel et al., 2008b), these data provide strong evidence that chondrite parent bodies, at least of the CO and CV chondrites, contain all components in a complementary relationship. That is, CAIs, AOAs, chondrules of various types, and matrix combine to produce a "chondritic" whole in major elements.

This "complementarity" provides perhaps the most powerful constraint on astrophysical models for planetesimal formation. It rules out multiple reservoir models in which chondrules formed in some (hot) place were then combined with cold matrix grains in a different place far away (thus preserving presolar grains and organics). Such combinations cannot magically produce the complementary relationships of composition reported here. Such models include the x-wind model of Shu et al. (2001), or turbulence-driven transport (Cuzzi et al., 2003) that require chondrules and refractory inclusions to originate near the sun, and to then combine with cold matrix dust far out in the disk. Complementarity also places severe constraints on models that form glassy beads from a panoply of possible colliding planetesimals, differentiated or not (e.g., Asphaug et al., 2011).

Models that are consistent with complementarity include those that form chondrules and CAIs from single gas+dust reservoirs with near-chondritic bulk compositions established prior to inclusion formation (Grossman, 1988). The essential mechanism must heat small regions to high (~1500 K) temperatures, while leaving nearby spaces cool (< 500 K for CO and < 800 K for CV; Huss et al., 2003; Mendybaev et al., 2002). It is increasingly clear that inclusion formation involved multiple reheating and complex exchange of isotopes of volatile elements (e.g., Ushikubo et al., 2013). Such a mechanism must affect various regions of the disk to different degrees of clast-forming efficiency. There are no inclusions in CI chondrites, and the ordinary chondrites (OC) are dominated by chondrules with essentially no CAIs or AOAs. The former experienced no inclusion formation, the latter highly efficient inclusion formation. Here, we have described characteristics of CO and CV chondrites, which are "in between" CI and OC.

## 6. CONCLUSIONS

Modern mapping techniques, coupled with 3D computed tomography to establish true clast sizes, are a way forward in understanding the accretion of meteorite parent bodies. Image analysis and data mining allow sufficient statistical power to quantify mineralogy, chemical composition, and textural relations pixel-by-pixel in each one of large numbers of inclusions in any particular chondrite.

We report inclusion/matrix ratios and inclusion modal abundances for a suite of CV and CO chondrites, and Acfer 094, by counting sufficient area and pixel density to provide representative statistics. These cover more area than previous studies and provide a self-consistent dataset for a large suite of C chondrites. We find nearly half the modal abundance of CAIs, and ~75% that of AOAs, reported by McSween (1977b) for CV. In CO chondrites, we find about the same CAI fraction, but only about 30% the AOA fraction as reported by McSween (1977a).

Acfer 094 (C2-ungr) has a lower inclusion/matrix ratio than all of the CO chondrites studied here, and much lower than the least equilibrated Colony CO 3.0 and Kainsaz CO 3.2 (Fig. 2). Texturally, the sizes of inclusions and their compositional variability in Acfer 094 are quite similar to those in the unequilibrated CO chondrites. Their fractionation of major elements between inclusions and matrix, normalized to modal abundance (Figs. 10, 12), is also similar.

Porosity, matrix abundance, and oxidative alteration are correlated in the CV chondrites. No such correlation exists for the CO chondrites. The porous, oxidized CV3 subgroup has nearly twice the volumetric matrix abundance, and nearly half the chondrule abundance, of the much less porous, reduced CV3 subgroup. The mean matrix grain abundance, less porosity, in the two subgroups is nearly identical. Others (e.g., Bunch and Chang, 1980) have proposed that the matrix brought with it the ice particles that promoted oxidative alteration of the CV parent body or bodies. Our correlation of matrix abundance, porosity and oxidative alteration lends strong support to this hypothesis.

Diameters of inclusions in CV are about five times larger than diameters of CO chondrite inclusions. Data in 3D will be needed to definitively show whether applying the Eisenhour (1996) 3D correction to apparent (2D) log-normal distributions of inclusion size results in a correct answer. Hypotheses of chondrule formation must address the constraints provided by these size differences.

Distribution of Fe, Mg and Si among inclusions and matrix in CO chondrites shows clearly the process of Fe-Mg exchange, with Si relatively immobile with increasing petrologic grade. No such trends are observed in the CV chondrites.

Element distributions among chondrules, refractory inclusions (CAI and AOA), and matrix provide further strong support for the complementary nature of chondrules and matrix in chondrites (Palme et al., 2015). Our data further suggest that CAIs and AOAs in these chondrites are also complementary, both to other CAIs and relative to chondrules and matrix. Inclusions formed at various efficiencies, making locally variably sized inclusions, which combined with local matrix into accreted bodies of chondritic composition from single reservoirs of chondritic (solar) composition.

## ACKNOWLEDGMENTS

The authors thank REU student Sarah McKnight (2009; Mount Holyoke, B.S. 2011) for her work on this project. We thank Charles Mandeville and Joe Boesenberg for technical help. Research was supported by U.S. N.A.S.A. grants NANG06GD89G and NNX10AI42G (DSE) and NNX12AI06G (MKW). The AMNH Physical Sciences REU Program (NSF #AST-055258, J. Webster and C. Liu PIs) supported H.R. (2006), C.E.B. (2007), K.L. (2008), and K.K. (2009). The AMNH HSSRP program supported M.L. and I.E. (2008-9). AMNH summer 2007 interns J. Finkelstein and G. Lutzky assisted in this project. This research has made use of NASA's Astrophysics Data System Bibliographic Services. The authors thank Drs. G. Libourel and A. Rubin for constructive and helpful reviews. Dr. A. Krot is thanked for thorough and swift editorial support.

## APPENDIX A. Supplementary data

Supplementary data associated with this article can be found, in the online version, at http://dx.doi.org/10.1016/j.gca.2015.10.007. An extended digital supplement containing x-ray maps, derived data, and software, is available through the AMNH Library at http://dx.doi.org/10.5531/sd.eps.2 .

**ELECTRONIC ANNEX**

Additional supporting information may be found in the online version of this article.

**TABLES**

Table S1: Identification scheme of McSween (1977a, b) for clast types.

*chondrules*: contain glass and/or have rounded external shapes (i.e., evidence they were once molten)
*inclusions*: irregularly shaped objects lacking unambiguous evidence of having once been liquid
*lithic & mineral fragments*: typically contain glass, and have mineralogy and internal textures identical to chondrules, hence considered to be derived from broken chondrules. Fragments of olivine throughout matrix are considered also to derive from the breakage of chondrules.

Within these three large divisions, four chondrule subtypes and two inclusion subtypes were identified:

*chondrule Type I*: This most abundant type contains granular forsterite ($Fa_{0-10}$), twinned clinoenstatite ($Fs_{0-7}$) and Ca-, Al-rich glass. Opaque minerals (magnetite, metal, sulfide) are frequently present as abundant disseminated grains. Type I chondrules are frequently concentrically structured, and opaque grains may concentrate at or near rims. McSween (1977c) reported abundances of opaque-mineral-rich and -poor subtypes of Type I chondrules.

*chondrule Type II*: Consist of porphyritic olivine ($Fa_{20-50}$) in brownish Ca-, Al-rich glass. These may be barred, and lack pyroxene or opaque phases in CO and CV chondrites.

*chondrule Type III*: Excentroradial pyroxene (i.e., RP of Lux et al. 19XX) chondrules contain fibrous pyroxene crystallites in glass, and are reported in CO but not in CV chondrites by McSween (1977b).

*chondrule Type IV*: Consist of melilite, anorthite, spinel, diopside and glass, similar to Ca-, Al-rich inclusions, but with a spherical shape. These are today commonly referred to as Type B CAIs.

*AOA inclusions*: Amoeboid olivine aggregates are irregularly shaped aggregates of clumps of aphanitic Ca-, Al-rich minerals (pyroxene, nepheline, perovskite, anorthite, etc.) surrounded by small anhedral olivine grains (Grossman and Steele, 1976). Spherical Type B CAIs (Type IV chondrules of McSween 1977b) may also be included in AOAs.

*refractory inclusions (CAI)*: These irregular clasts are rich in refractory Ca, Al and Ti, in minerals such as melilite, spinel, perovskite, clinopyroxene, anorthite, and less often hibonite, cordierite, olivine, grossite, with nepheline, sodalite and grossular interpreted to be secondary alteration minerals. Grain sizes range from coarse to very fine grained, but less coarse than chondrules. Mineral species may appear to be layered.

Table S2: Additional information on x-ray mapping.

| Meteorite: | museum id. | WDS | EDS (*=WDS) | A(cm2) |
|---|---|---|---|---|
| Allende | 5046-1A | (1) Si,Fe,Ca,Mg,Al | S, Ti, Ni | 18.0875 |
| Allende | 5046-1C | (1) Si,Fe,Ca,Mg,Al | S, Ti, Ni | 11.2546 |
| Allende | 4884-s1B | (1) Si,Fe,Ca,Mg,Al | S, Ti, Ni | 12.5724 |
| Allende | 4884-s2B | (1) Si,Fe,Ca,Mg,Al | S, Ti, Ni | 15.6798 |
| Allende | 4884-s2B-F1 | (1) Si,Fe,Ca,Mg,Al | Na, Ti, S | 0.0584 |
| Allende | 4884-s2B-F2 | (1) Si,Fe,Ca,Mg,Al | Na, Ti, S | 0.0564 |
| Allende (AL2ps2) | 4884-t2-ps2A | Si,Fe,Ti,Mg,Ca 10ms | *Al,Ni,Mn,Na,S 15ms | 0.3271 |
| Allende (AL2ps5) | 4884-t2-ps5A | (2) Si,Fe,Ti,Mg,Ca | *Al,Ni,K,Na,S | 0.8981 |
| Allende (AL2ps9) | 4884-t2-ps9A | (2) Si,Fe,Ti,Mg,Ca | *Al,Ni,Mn,Na,S | 0.8396 |
| Allende | 4288-1 | (1) Si,Fe,Ca,Mg,Al | S, Ti, Ni | 2.1649 |
| Tibooburra | 5003-1 | (1) Na,Fe,Ca,Mg,Al | S, Si, Ti | 2.8962 |
| Nova 002 | 4826-2 | (1) Si,Fe,Ca,Mg,Al | S, Ti, Ni | 0.9395 |
| Bali | 4936-1 | (1) Si,Fe,Ca,Mg,Al | S, Ti, Mn | 1.9398 |
| Mokoia | 3906-4 | (1) Si,Fe,Ca,Mg,Al | S, Ti, Mn | 0.8389 |
| Vigarano | 2226-4 | (1) Si,Fe,Ca,Mg,Al | S, Ti, Ni | 0.9854 |
| Leoville | USNM-3535-1 | (2) Al,Ni,Ti,Mg,Ca | S, Si, Fe | 1.4832 |
| | | | total: | **71.022** |
| Meteorite: | sample | WDS | EDS (*=WDS) | A(cm2) |
| Colony | 4595-1 | (2) Mg,Ni,Ti,Al,Ca | S, Si, Fe | 0.3252 |
| Kainsaz | 4717-1-1Cp1 | (2) Mg,Ni,Ti,Al,Ca | S, Si, Fe | 0.2363 |
| Kainsaz | 4717-1-1Cp2 | (2) Mg,Ni,Ti,Al,Ca | S, Si, Fe | 0.2241 |
| Ornans | 520-1-r4 | (2) Mg,Ni,Ti,Al,Ca | S, Si, Fe | 0.1402 |
| Lance | 618-1 | (2) Al,Ni,Ti,Mg,Ca | S, Si, Fe | 0.4589 |
| Warrenton | 4151-1 | (2) Mg,Ni,Ti,Al,Ca | S, Si, Fe | 0.1447 |
| Acfer 094 | IfP-PL93022 | (2) Mg,Ni,Ti,Al,Ca | S, Si, Fe | 0.1045 |
| | | | total: | **1.634** |

**FIGURES**

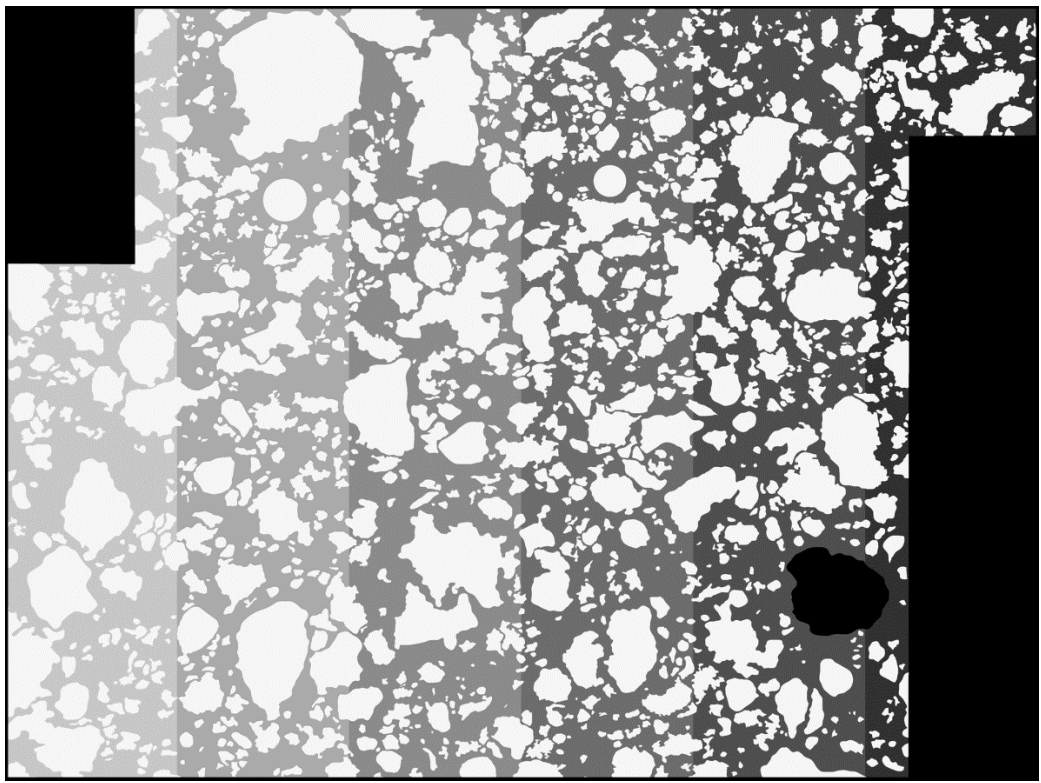

Fig S1: Measurement of matrix heterogeneity in Acfer 094 map. Mean element counts/pixel were tabulated for 9 evenly spaced vertical slices of the mapped area. The matrix portion of each slice has an increasing grayscale from the leftmost to rightmost slice. All inclusions are white (cf., Fig. 1f). The fraction of mapped area (black is masked off) that is matrix is $55 \pm 5\%$ over the 9 areas.

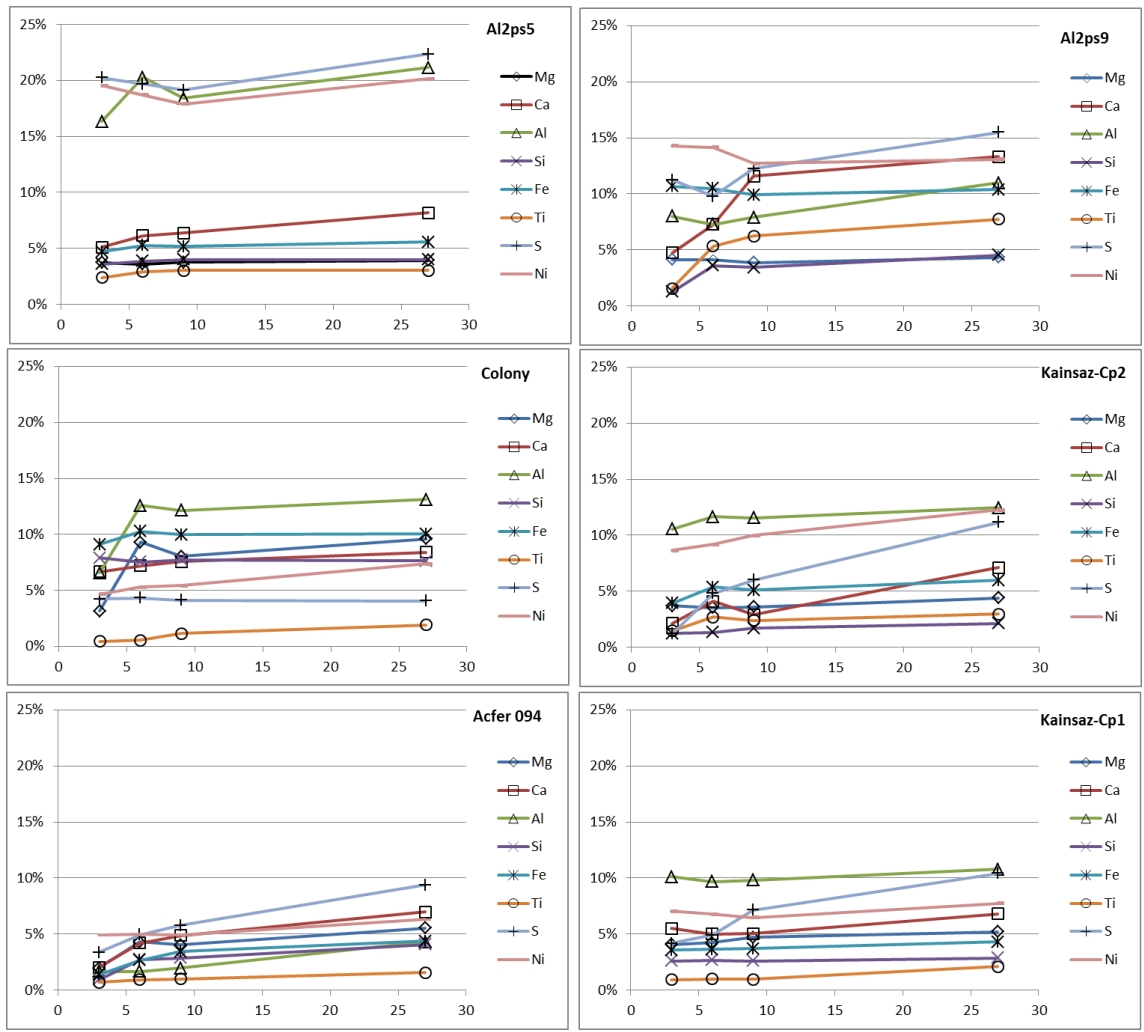

Fig S2: Heterogeneity of matrix composition. Standard deviation (1σ) of element counts/pixel in matrix only, among 3, 6, 9, and 27 slices of mapped areas.

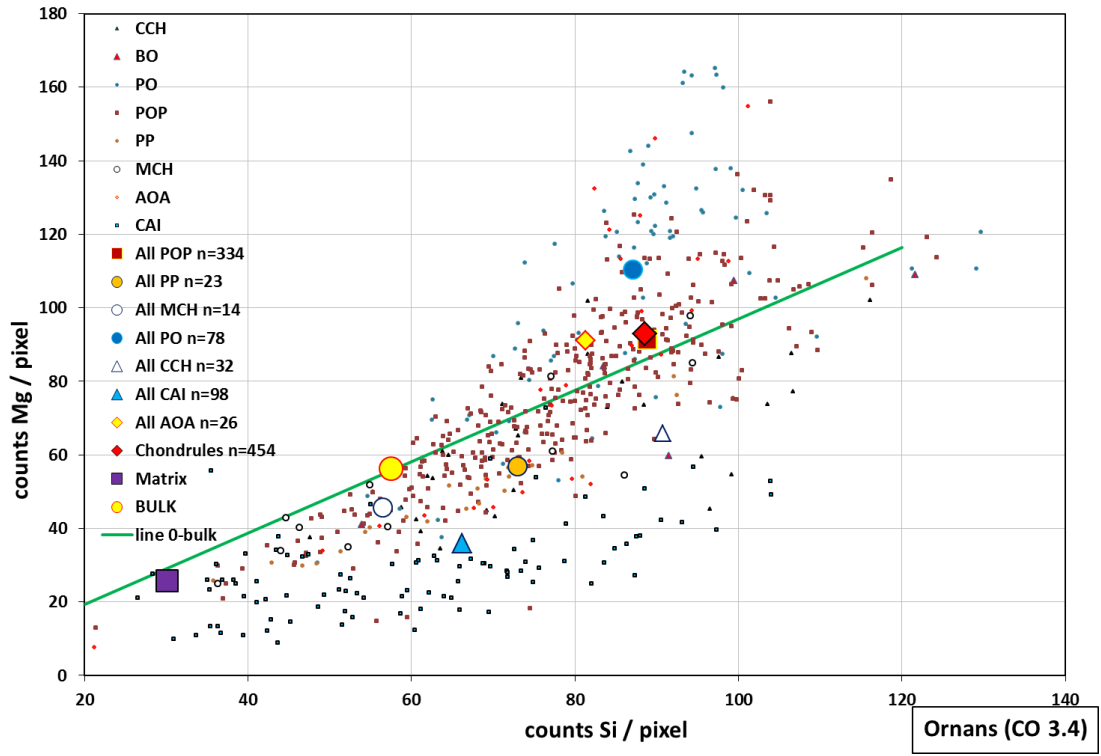

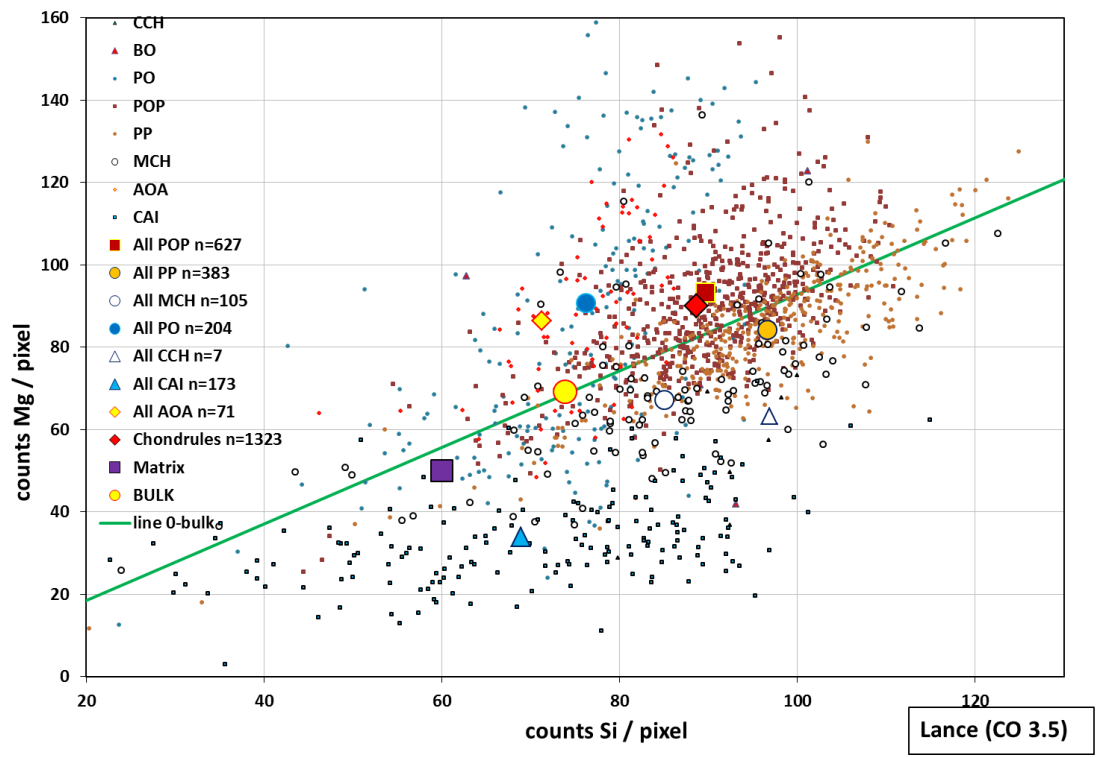

Fig S3: Per pixel counts of Si and Mg for all clasts, mean clasts, matrix, and bulk meteorite for Ornans (CO 3.4) and Lancé (CO 3.5). Solid green line joins bulk ratio to origin.

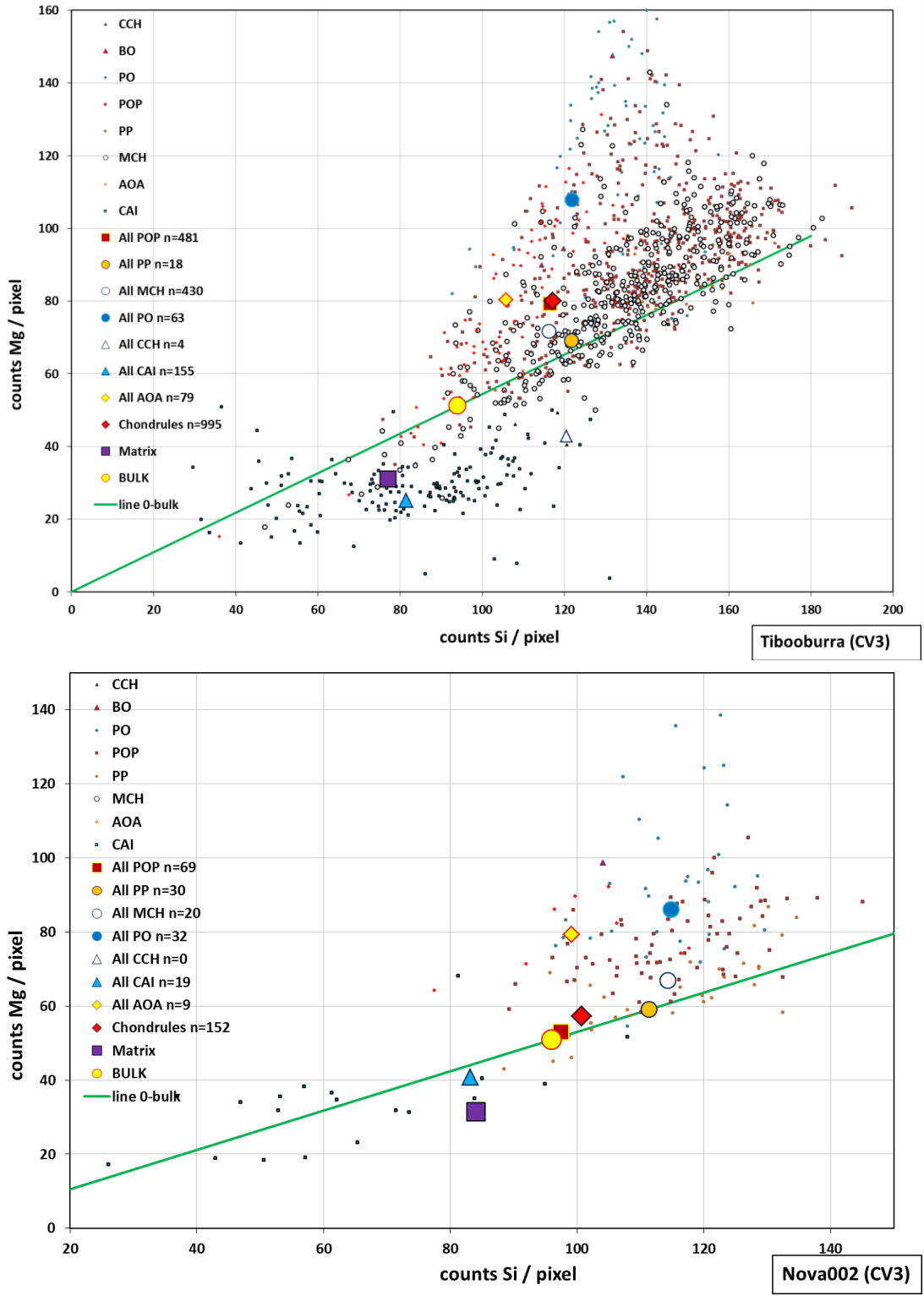

Fig. S4: Per pixel counts of Mg and Si for Tibooburra (CV3-Ox) and Nova 002 (CV3-R). Solid green line joins bulk ratio to origin.

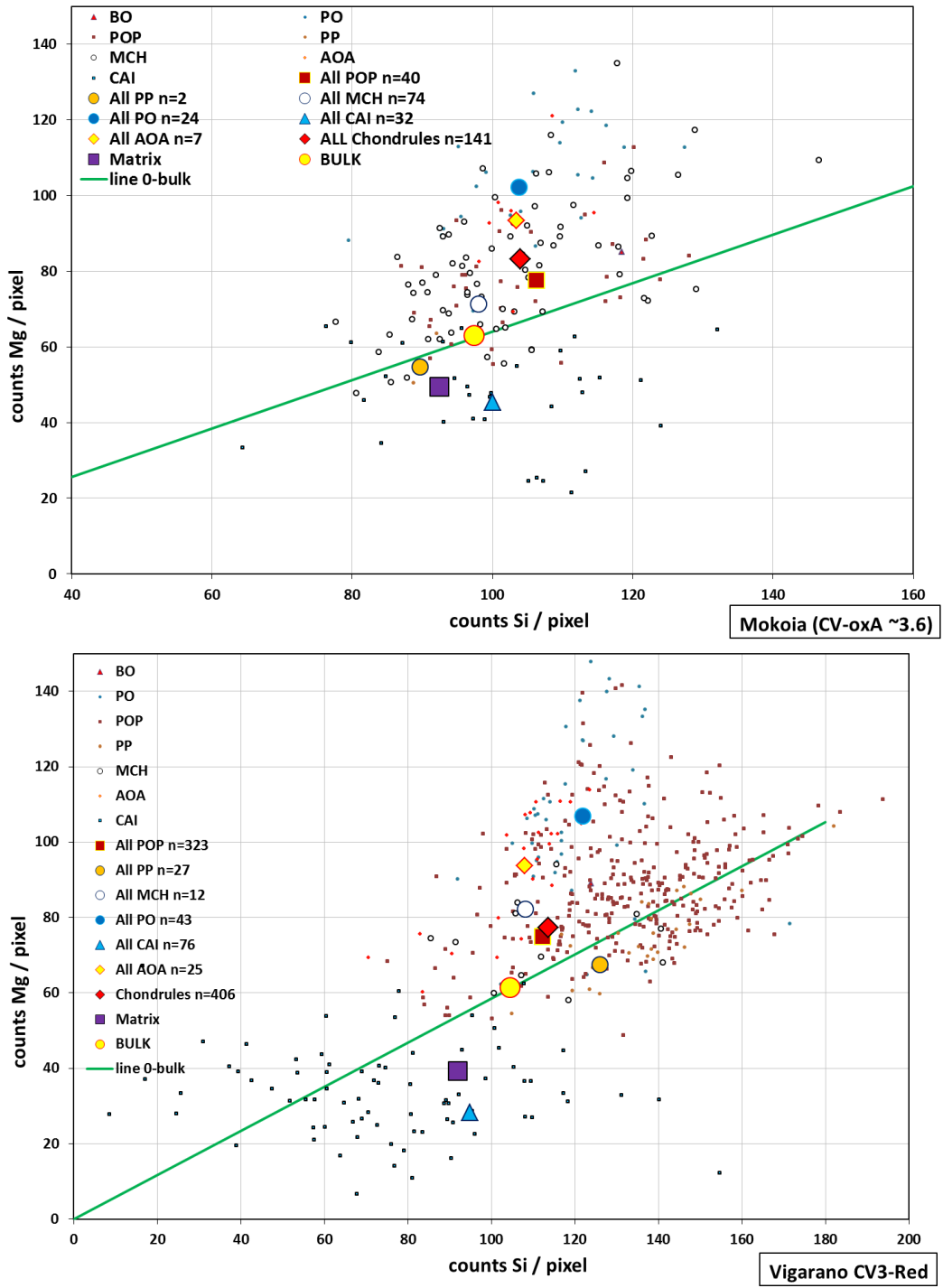

Fig. S5: Per pixel counts of Mg and Si for all clasts, mean clasts, matrix, and bulk meteorite for Mokoia (CV-OxA) and Vigarano (CV-R). Solid green line joins bulk ratio to origin.

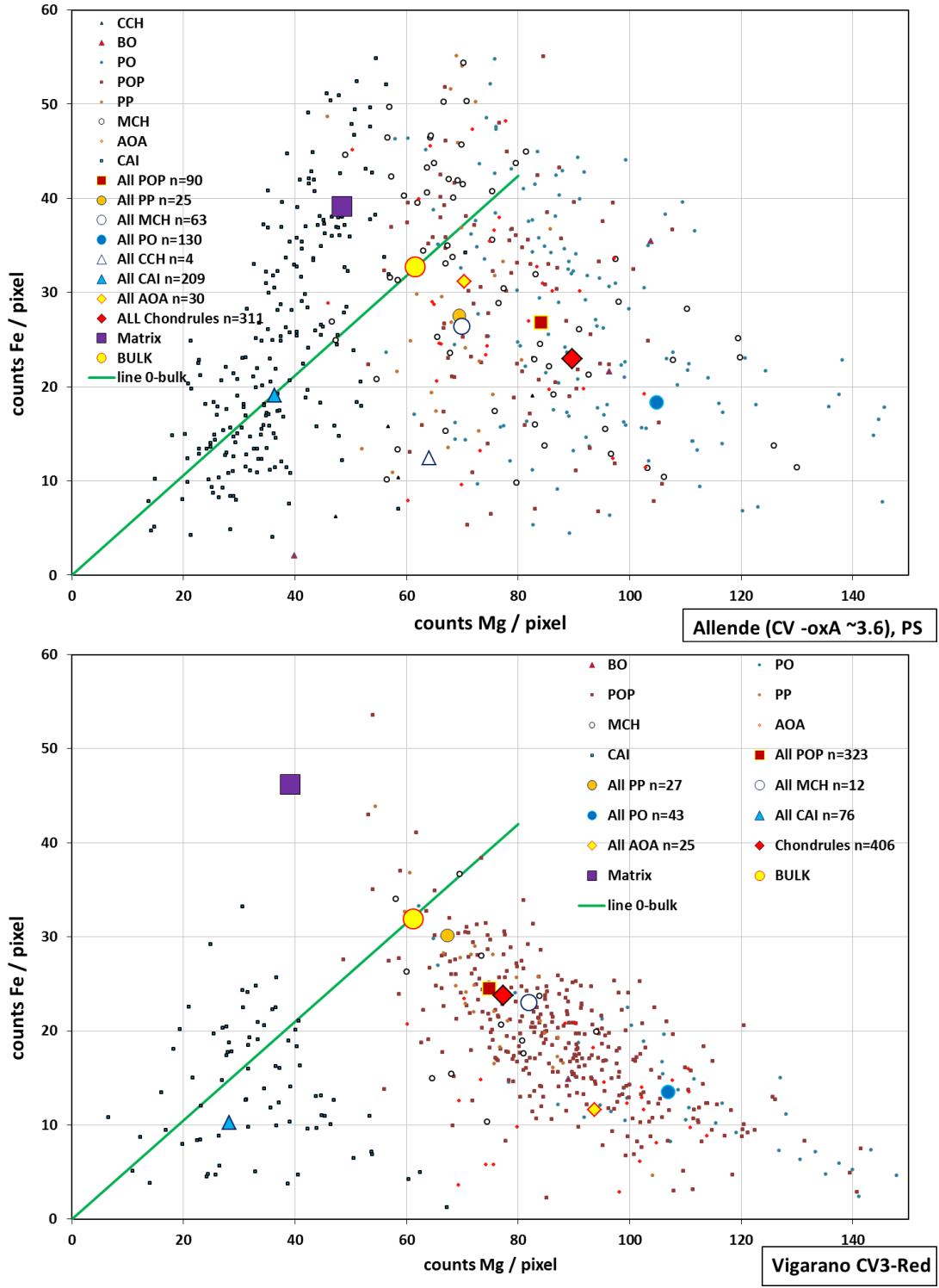

Fig. S6: Per pixel counts of Mg and Fe for all clasts, mean clasts, matrix, and bulk meteorite for two Allende PS (CV-OxA) and Vigarano (CV-R). Solid green line joins bulk ratio to origin.

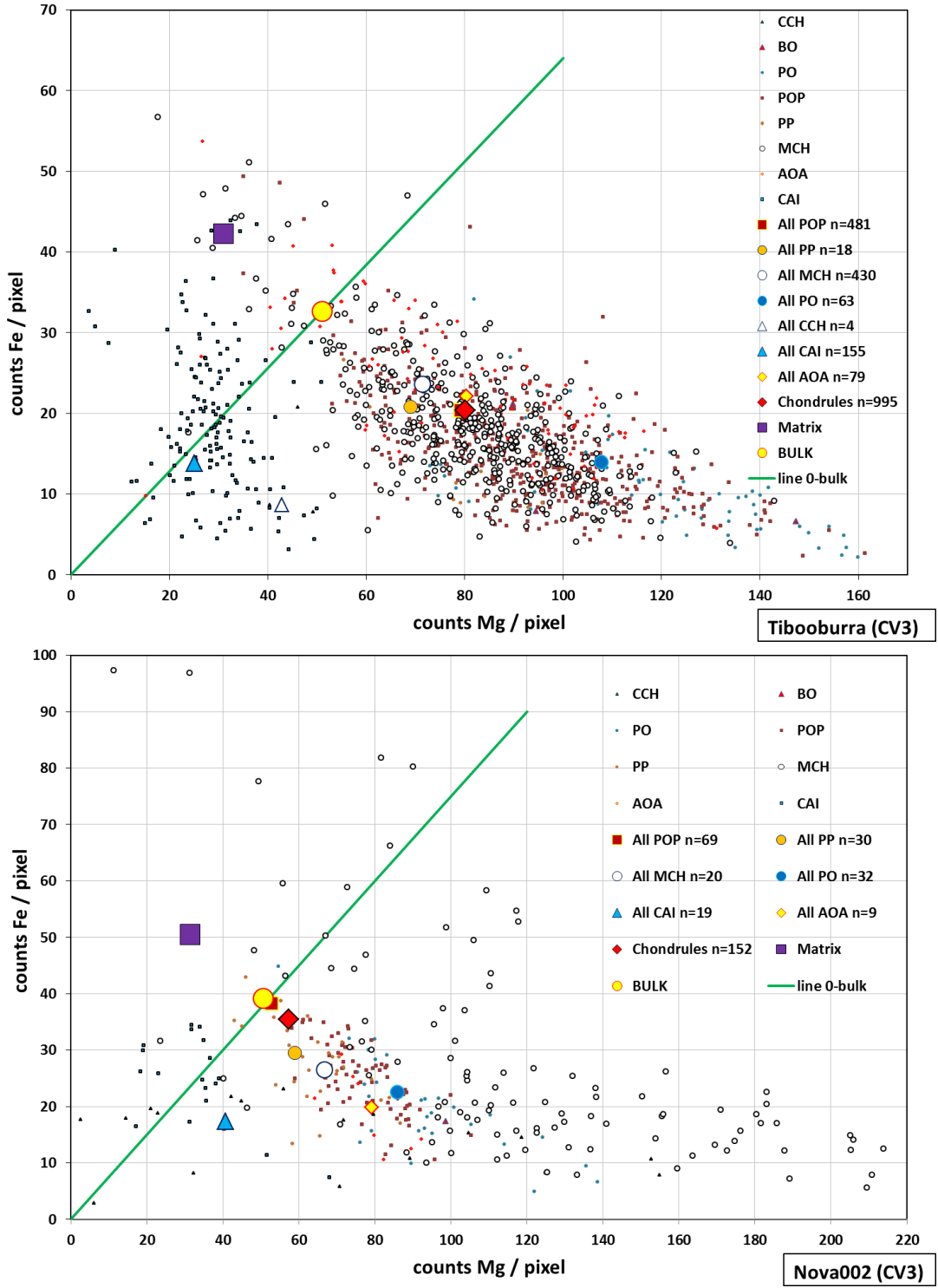

Fig. S7: Per pixel counts of Mg and Fe for Tibooburra (CV3-Ox) and Nova 002 (CV3-R). Solid green line joins bulk ratio to origin.

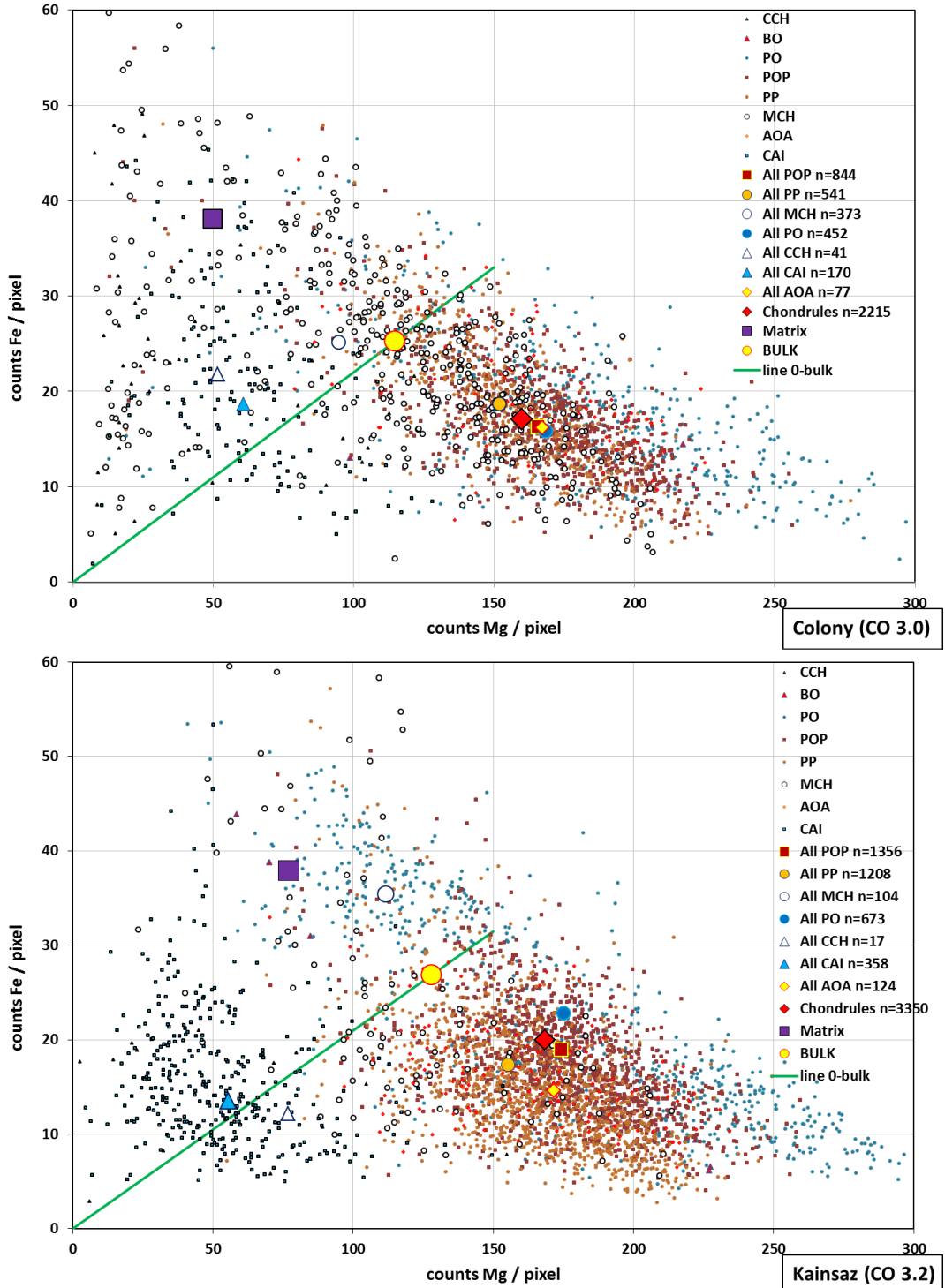

Fig S8: Per pixel counts of Mg and Fe for clasts, mean clasts, matrix, and bulk meteorite for Colony (CO 3.0) and Kainsaz (CO 3.2). Solid green line joins bulk ratio to origin.

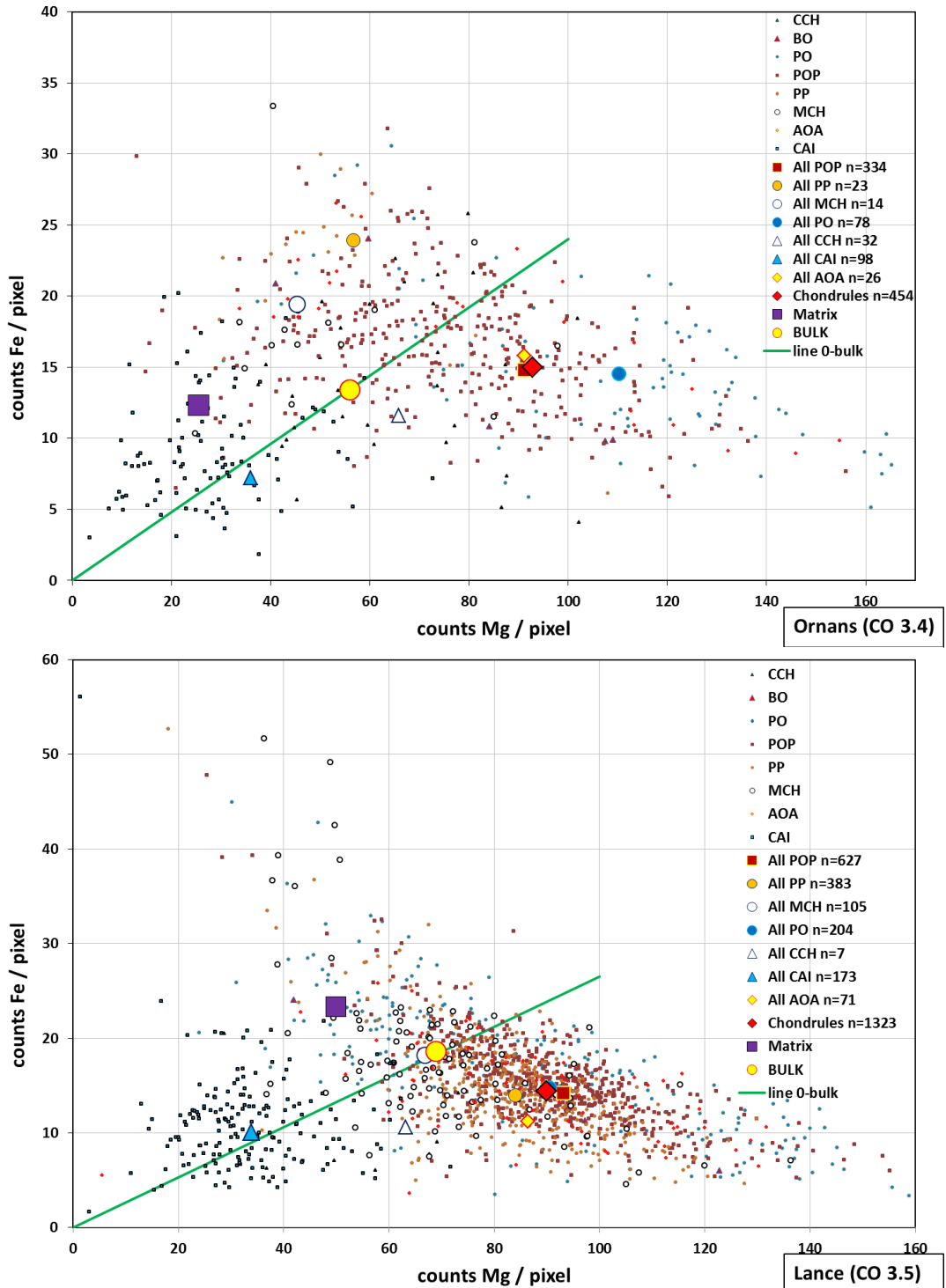

Fig. S9: Per pixel counts of Mg and Fe for all clasts, mean clasts, matrix, and bulk meteorite for Ornans (CO 3.4) and Lancé (CO 3.5). Solid green line joins bulk ratio to origin.

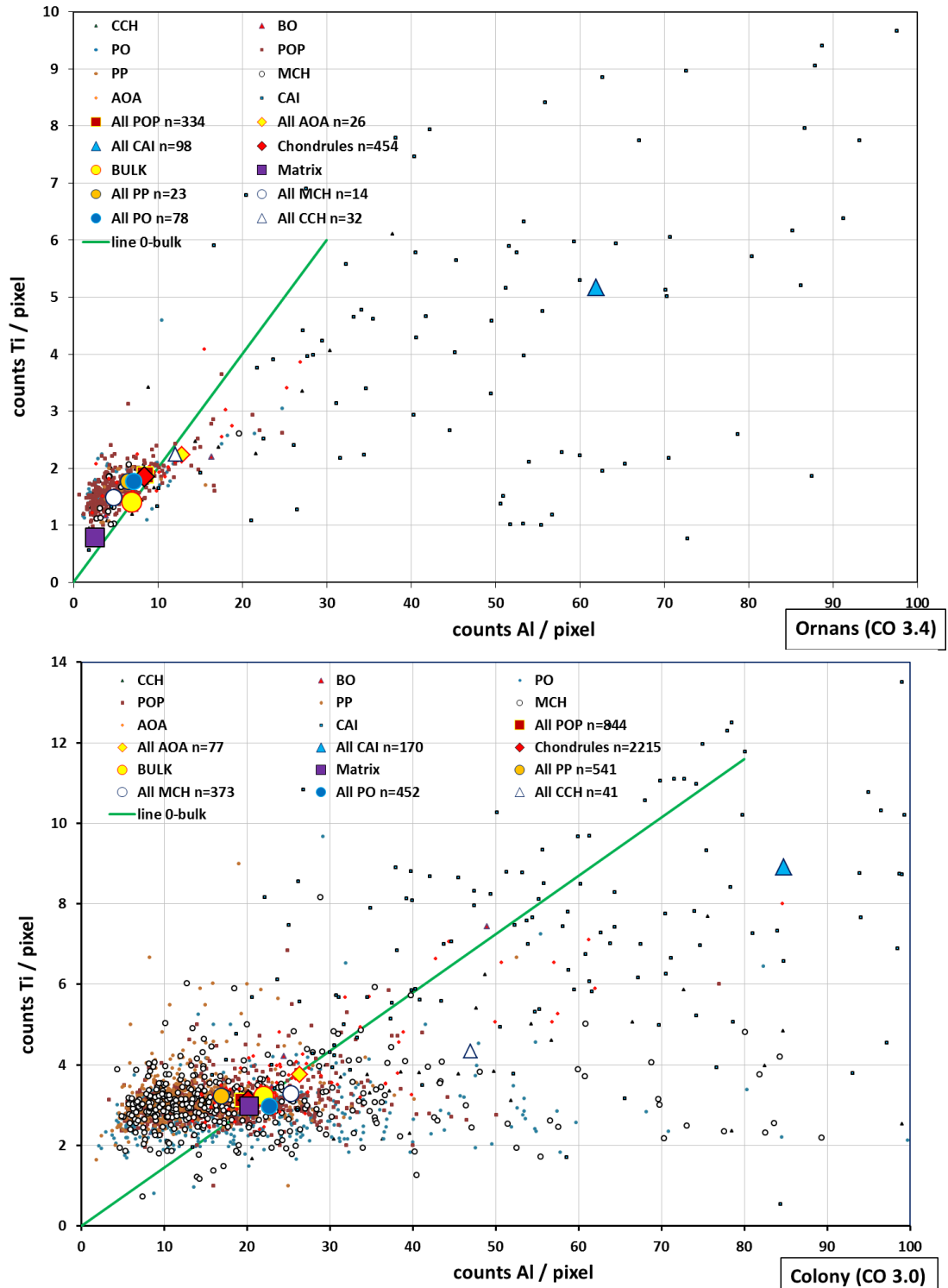

Fig. S10: Per pixel counts of Ti and Al for Ornans (CO 3.4) , and Colony (CO 3.0). Solid green line joins bulk ratio to origin.

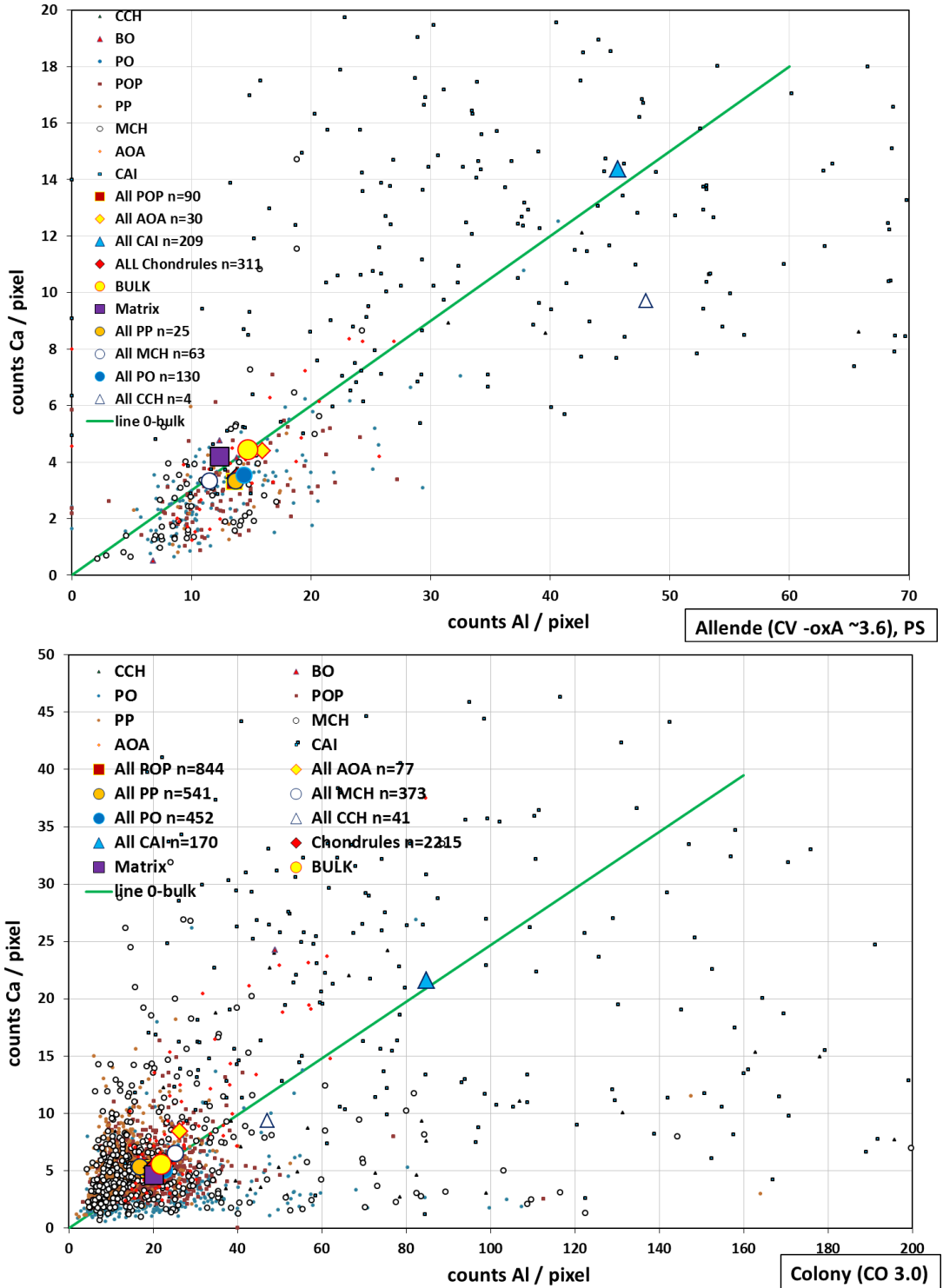

Fig. S11: Per pixel counts of Al vs Ca, for clasts, mean clasts, matrix, and bulk meteorite for Allende (CV-OxA ~3.6) and Colony (CO 3.0). Solid green line joins bulk ratio to origin.

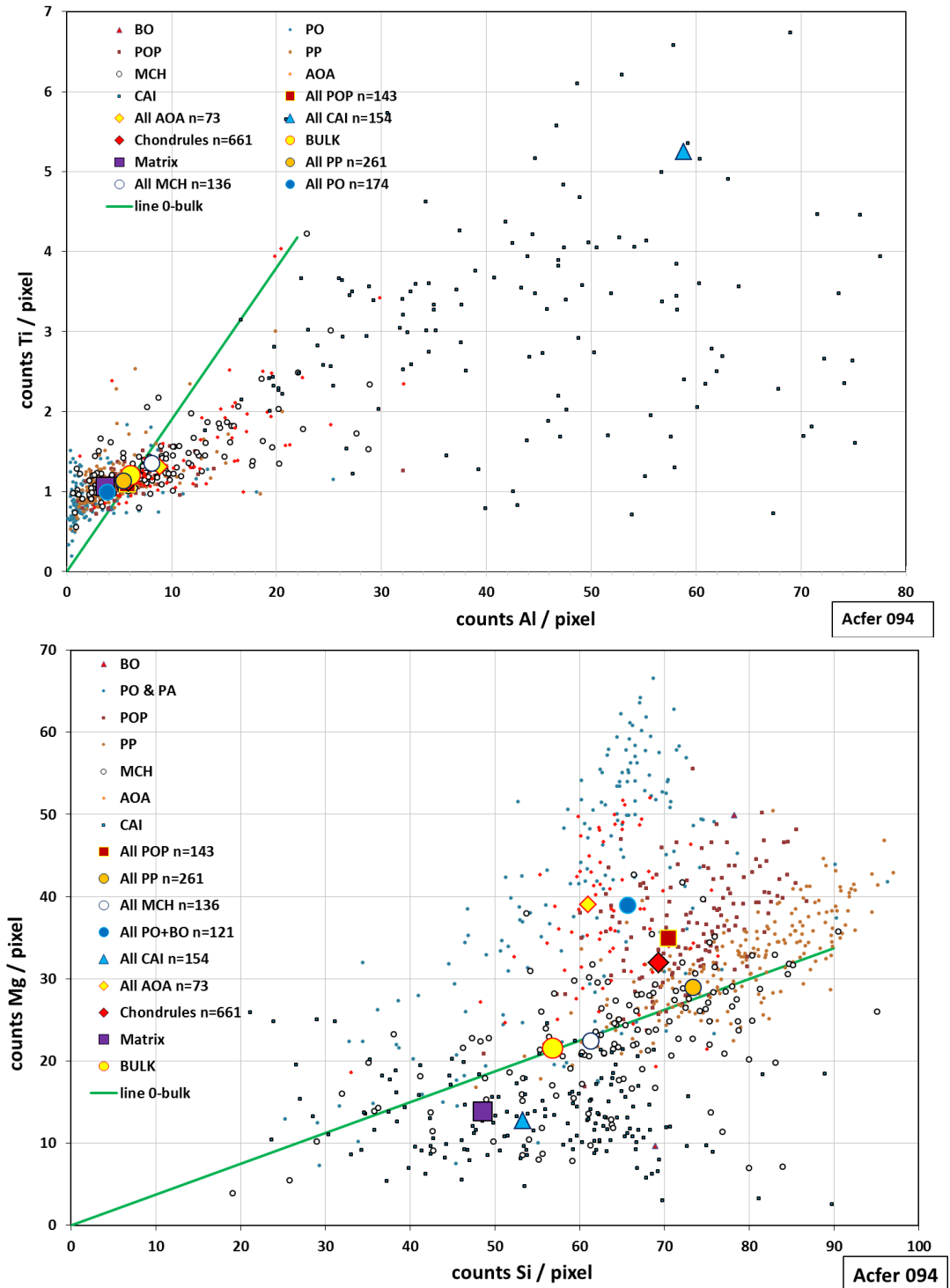

Fig. S12: Compositions of Ti vs Al and Mg vs Si in inclusions in Acfer 094 (C2-ungr). Solid green line joins bulk ratio to origin.